\documentclass[longauth, traditabstract]{aa}

\newcommand{\Av}{{A$_V$}}
\newcommand{\cmsq}{{cm$^{-2}$}}
\newcommand{\cmc}{{cm$^{-3}$}}
\newcommand{\mstar}{M$_\odot$}
\newcommand{\IRAS}{{\it IRAS}}

\newcommand{\Planck}{{\it Planck}}
\newcommand{\spitzer}{{\it Spitzer}}
\newcommand{\lum}{L$_\odot$} 
\newcommand{\hubble}{{\it Hubble}}
\newcommand{\microm}{$\mu$m}
\newcommand{\um}{$\mu$m}

\newcommand{\disperse}{DisPerSE}

\newcommand{\hii}{{H{\scriptsize II}}}
\newcommand{\herschel}{{\it Herschel}}

\usepackage{natbib}
\bibpunct{(}{)}{;}{a}{}{,} \usepackage{graphicx, pstricks}
\usepackage{wasysym,afterpage}
\usepackage{rotating}
\usepackage{txfonts}

\begin{document}
   \title{The M16 molecular complex under the influence of NGC\,6611. \herschel\thanks{\herschel\ is an ESA space observatory with science instruments provided by European-led Principal Investigator consortia and with important participation from NASA.}'s perspective of the heating effect on the Eagle Nebula\thanks{Figs. \ref{fig:pdf} \& \ref{fig:rp:all} are only available in electronic form at http://www.aanda.org.}}
\titlerunning{\herschel's perspective of heating on the Eagle Nebula}
\authorrunning{T. Hill et al.}
        \author{T. Hill\inst{1}
          \and
          F. Motte\inst{1}
          \and
          P. Didelon\inst{1}
          \and
          G.~J.~White\inst{2,3}
          \and
          A. P. Marston\inst{4}
          \and
          Q. Nguy$\tilde{\hat{\rm e}}$n Lu{\hskip-0.65mm\small'{}\hskip-0.5mm}o{\hskip-0.65mm\small'{}\hskip-0.5mm}ng\inst{1}
          \and
          S. Bontemps\inst{5}
          \and
          Ph. Andr\'e \inst{1}
          \and
          N.~Schneider\inst{1}
          \and
          M. Hennemann\inst{1}
          \and
          M. Sauvage\inst{1}
          \and
          J. Di Francesco\inst{6}
          \and
          V.~Minier\inst{1}
          \and
          L. D. Anderson\inst{7}           \and
          J.~P. Bernard\inst{8}           \and
          D. Elia\inst{9}
          \and
          M. J. Griffin\inst{10}
          \and
          J.~Z.~Li           \inst{11}
          \and
          N.~Peretto\inst{1}
          \and
          S.~Pezzuto           \inst{9}
          \and
          D.~Polychroni           \inst{9}
          \and
          H.~Roussel           \inst{12}
          \and
          K.~L.~J.~Rygl           \inst{9}
          \and
          E. Schisano\inst{9}          \and
          T.~Sousbie\inst{12}
          \and
          L.~Testi\inst{13}           \and
          D.~Ward~Thompson\inst{10,14}           \and
          A. Zavagno\inst{15}
                                                }
   \institute{Laboratoire AIM, CEA/IRFU CNRS/INSU Universit\'e Paris Diderot, CEA-Saclay, 91191 Gif-sur-Yvette Cedex, France\\
              \email{tracey.hill@cea.fr}
         \and          The Rutherford Appleton Laboratory, Chilton, Didcot, OX11 0NL, UK
         \and          Department of Physics and Astronomy, The Open University, Milton Keynes, UK
         \and          Herschel Science Centre, ESAC, Spain
         \and          Universit\'e de Bordeaux, OASU, Bordeaux, France
         \and          National Research Council of Canada, 5071 West Saanich Road, Victoria, BC, Canada V9E 2E7.
         \and          Department of Physics, West Virginia University, Morgantown, WV 26506, USA
         \and          Universit\'e de Toulouse, UPS, CESR, 9 avenue du colonel Roche, 31028 Toulouse Cedex 4, France; CNRS, UMR5187, 31028 Toulouse
         \and          IAPS- Instituto di Astrofisica e Planetologia Spaziali, via Fosso del Cavaliere 100, 00133 Roma, Italy
         \and          School of Physics and Astronomy, Cardiff University, Queens Buildings, The Parade, Cardiff, CF243AA, UK
         \and          National Astronomical Observatories, Chinese Academy of Sciences, A20 Datun Road, Chaoyang District, Beijing 100012, China
         \and          Institut d'Astrophysique de Paris, Universit\'e Pierre et Marie Curie (UPMC), CNRS (UMR 7095), 75014 Paris, France 
         \and          ESO, Karl Schwarzschild str. 2, 85748 Garching bei Munchen, Germany.
         \and          Jeremiah Horrocks Institute, University of Central Lancashire, PR1 2HE, UK 
         \and          Laboratoire d'Astrophysique de Marseille UMR6110, CNRS, Universit\'e de Provence, 38 rue F. Joliot-Curie, 13388 Marseille, France\\
                  \thanks{}
             }

   \date{February 2012}

 \abstract{We present \herschel\ images from the HOBYS key program of the Eagle Nebula (M16) in the far-infrared and sub-millimetre, using the PACS and SPIRE cameras at 70\,\um, 160\,\um, 250\,\um, 350\,\um, 500\,\um. M16, home to the Pillars of Creation, is largely under the influence of the nearby NGC\,6611 high-mass star cluster.  The \herschel\ images reveal a clear dust temperature gradient running away from the centre of the cavity carved by the OB cluster.
We investigate the heating effect of NGC\,6611  on the entire M16 star-forming complex seen by \herschel\ 
including the diffuse cloud environment \emph{and} the dense filamentary structures identified in this region. In addition, we
interpret the three-dimensional geometry of M16 with respect to the nebula, its surrounding environment, and the NGC\,6611 cavity. 
The dust temperature and column density maps reveal a prominent eastern filament running north-south and away from the high-mass star-forming central region and the NGC\,6611 cluster, as well as a northern filament which extends around and away from the cluster across a potential ridge of star formation. The dust temperature in each of these filaments decreases with increasing distance from the NGC\,6611 cluster, indicating a heating penetration depth  of $\sim$\,10\,pc in each direction in 3\,--\,6\,$\times$\,10$^{22}$\,\cmsq\ column density filaments.  We show that in high-mass star-forming regions OB clusters impact the temperature of future star-forming sites, modifying the initial conditions for collapse and effecting  the evolutionary criteria of protostars developed from spectral energy distributions.
 Possible scenarios for the origin of the morphology seen in this region are discussed, including a western equivalent to the eastern filament, which was destroyed by the creation of the OB cluster and its subsequent winds and radiation. 
}

   \keywords{ISM: individual objects (M16, NGC\,6611) --
     ISM: filaments --
     ISM: structure --
     ISM: dust, extinction --
     Stars: early-type --
          Stars:~protostars      
           }

   \maketitle
\section{Introduction}\label{sec:intro}

Forming in dynamic and ever-changing environments, high-mass stars are complex entities which, through radiation and stellar winds, heavily impact their natal environments, possibly even triggering the formation of new stars and planetary systems. A general rarity of candidates and their rapid evolutions mean that it is difficult to characterise the early evolutionary stages of high-mass stars \citep[cf.][]{garay99}.  

\begin{figure*}
\includegraphics[height=0.5\textwidth]{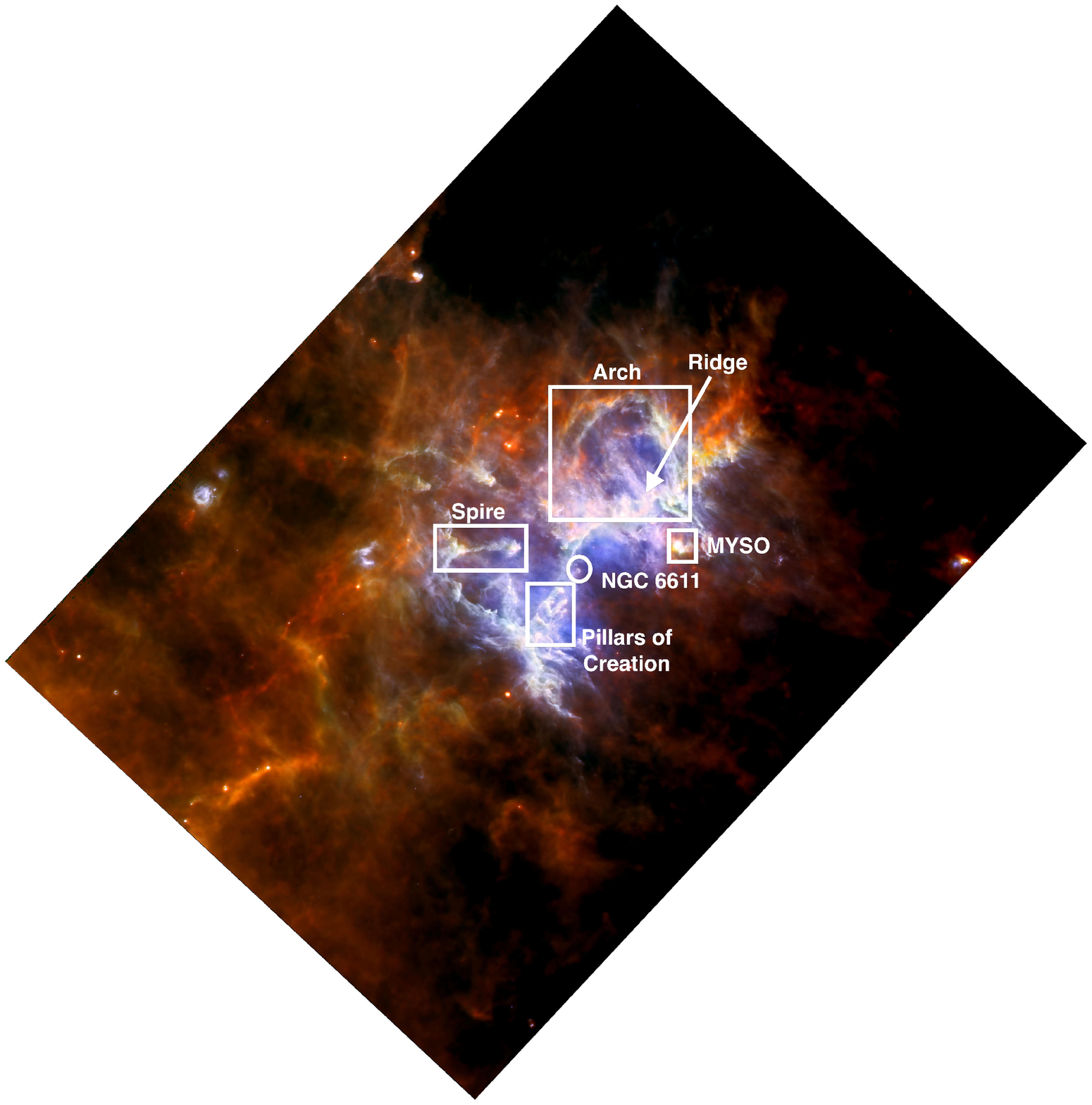}
\begin{pspicture}(0,0)
\psline[linecolor=white, linewidth=1pt](-5.3,0.4)(-0.5,5.5)
\rput[linecolor=cyan](-3.7,3.0){\white 30\,pc}
\end{pspicture}
\hspace{0.4cm}
\includegraphics[height=0.45\textwidth]{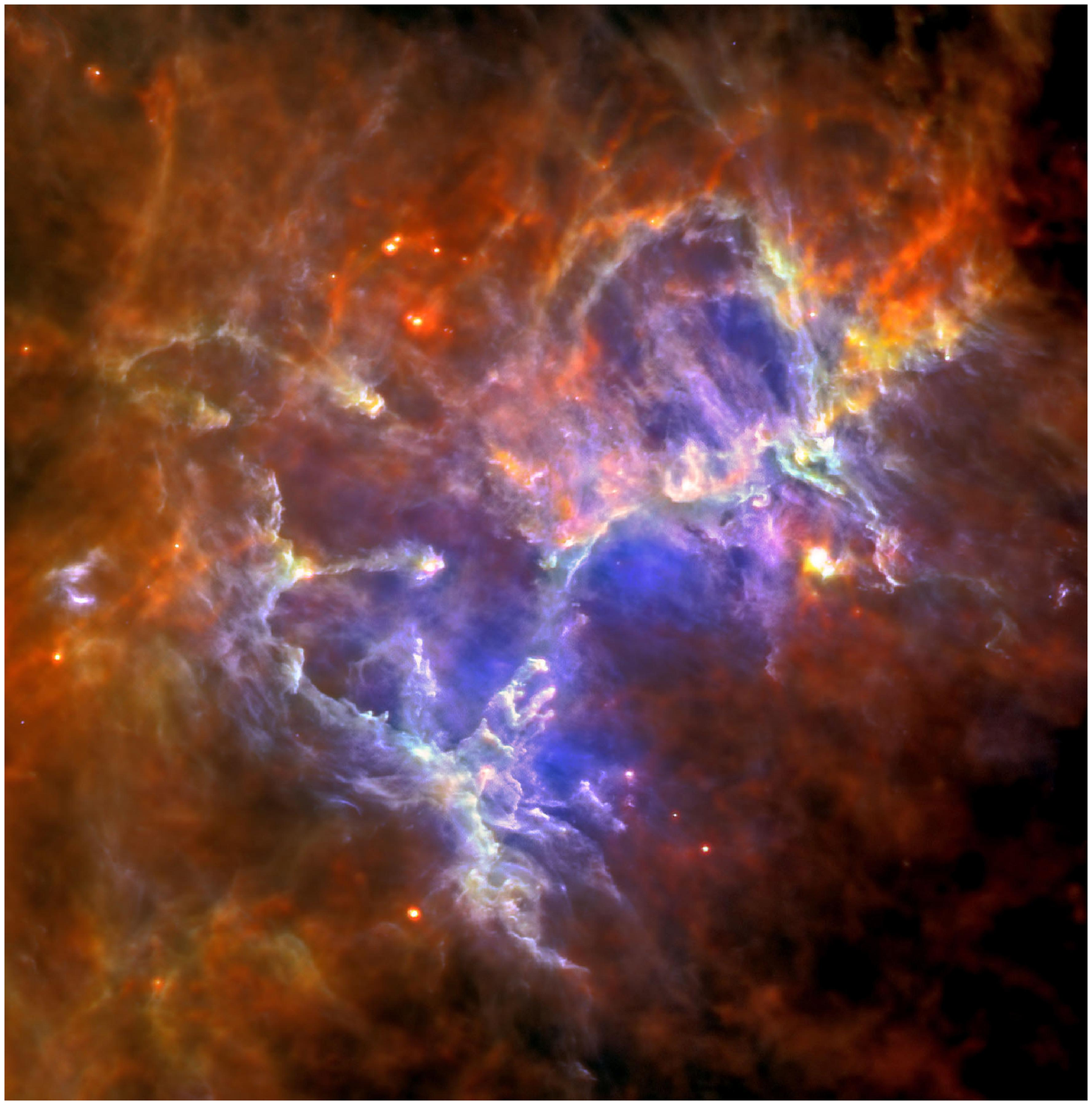}
\begin{pspicture}(0,0)
\psline[linecolor=white, linewidth=1pt](-7.8,0.5)(-0.5,0.5)
\rput[linecolor=cyan](-4.5,0.7){\white 15\,pc}
\end{pspicture}
\caption{Three-colour image of M16 using 70\,\um (blue), 160\,\um (green), 250\,\um (red). Left:
Image of the entire M16 complex with well-known regions from the literature, as well as regions identified (the arch and ridge features) and discussed in this paper, indicated on the map. Right: Zoom to the central region of M16 (see left image).   \label{fig:hipe}}
\end{figure*}

The \herschel\ Space Observatory \citep[launched 2009;][]{pilbratt10} has already had a significant impact on star formation studies, owing primarily to the large key programs devoted to this field, specifically the \herschel\ imaging study of OB Young Stellar Objects \citep[HOBYS;][]{motte10} and Gould Belt Survey \citep{andre10}, as well as the Hi-Gal survey \citep{molinari10}.  These programs target the natal environments of burgeoning young stars, such as interstellar filaments and ridges \citep{arzoum11, hill11}, as well as sources embedded within them, e.g., prestellar cores and protostars \citep{bontemps10, hennemann10, giannini12}, and investigate the impact of OB stars on cloud formation \citep{zavagno10, schneider10}. The primary HOBYS\footnote{http://www.herschel.fr/cea/hobys/en/} driver is to compile an unbiased census of high-mass star progenitors, and their natal ridges, in the nearest ($<$\,3\,kpc) star-forming complexes \citep[][]{motte10}, one of which is M16.

The Eagle Nebula (M16) is a young \citep[1\,--\,3\,$\times$\,10$^6$\,yr;][]{hillenbrand93}, active high-mass star-forming region in the constellation of Serpens $\sim$\,2\,kpc from the Sun.  M16 was first observed by de Cheseaux in 1745 and then rediscovered by Messier in 1764 who was the first to comment on its nebulosity. The European Space Agency's (ESA's) Infrared Space Observatory \citep[ISO;][]{cesarsky96} revealed the presence of unusually hot dust in the core of the nebula \citep[][T$_{colour}$~$\sim$\,200\,K]{pilbratt98}.  But it was the \hubble\ Space Telescope images of its ionised gas \citep{hester96}, better known as the `Pillars of Creation', that has made M16  iconic and one of the most observed objects in the sky.

M16, and especially these pillars, has since been observed at many wavelengths spanning the electromagnetic spectrum, as well as in many dedicated spectral line studies \citep[e.g.,][to name a few]{pound98,white99, allen99, urq03,linsky07}. More recently, M16 was observed in the mid-infrared with IRAC (3.6--8.0\,\microm) as part of the \spitzer\ GLIMPSE legacy program \citep[][]{indebetouw07}, and in the far-infrared with MIPS (24\,\um\ \& 70\,\microm) as part of the \spitzer\ MIPSGAL program \citep[][]{flagey11}. The former survey targeted the warm dust toward this region and revealed its (large) population of young stellar objects (YSOs) that these authors deemed protostellar candidates. The latter paper examined the dust emission and composition, such as the grain type and size, in this region, and concluded that a supernova remnant may be responsible for heating the inner portion of M16.

Adjacent to the Eagle Nebula, and responsible for heating and ionising the Nebula, is the young open star cluster NGC\,6611. At a distance of $\sim$~1.75\,--\,2\,kpc \citep{gva08}, NGC\,6611 contains four early-type O stars \citep[$\sim$\,\,2\,--\,3\,Myr;][]{dufton06}, and is also associated with the cluster of a B2.5I star \citep[$\sim$\,6\,Myr;][]{hillenbrand93}. The total stellar mass of the NGC\,6611 cluster is $\sim$\,2$\,\times$\,10$^4$\,\mstar\ \citep{wolff07}. See \citet{gva08} and \citet{belikov00} for more information on the NGC\,6611 cluster.

We present here unprecedented high-resolution high-sensitivity \herschel\ observations, spanning the far-infrared and submillimetre regime, of the Eagle Nebula. These \herschel\ data provide unique access to this far-infrared regime at much greater spatial resolution, and additionally cover a greater area around the central NGC\,6611 cluster and the Pillars of Creation, than previous (space) observations and studies. \herschel\ allows us to probe the cold component of the dust, complementary to \spitzer, and to deepen the study of warm dust (through the 70\,\um\ band) in this region.
Here, we present a structural analysis of the \emph{entire} M16 molecular cloud complex, with respect to the well-known central OB cluster. A companion paper by \citet{white12} examines the protostellar content of the M16 region as detected by \herschel. 
\begin{figure*}
\hspace{-2mm}
\includegraphics[height=0.45\textwidth]{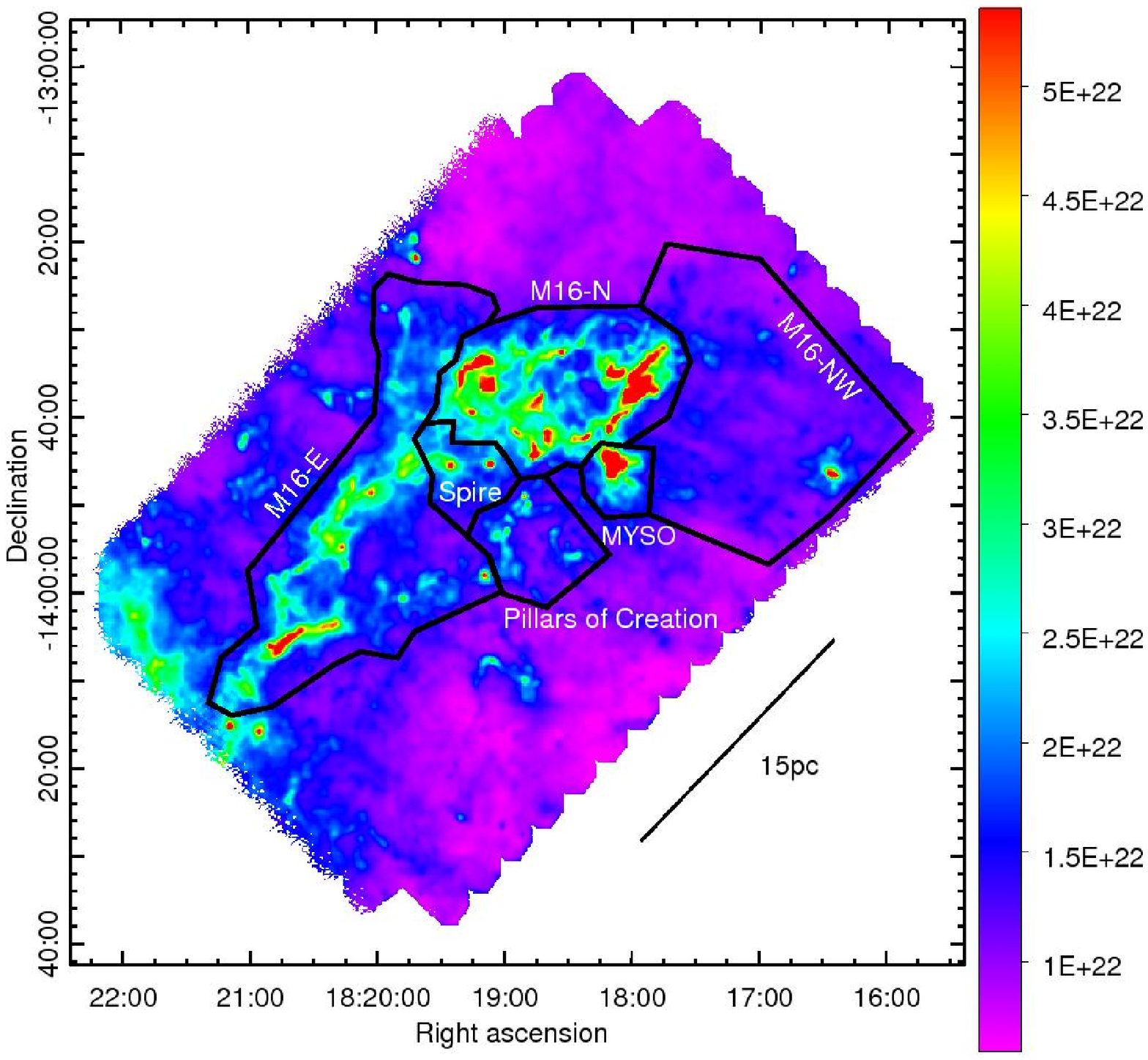}
 \hspace{-5mm}
\includegraphics[height=0.48\textwidth]{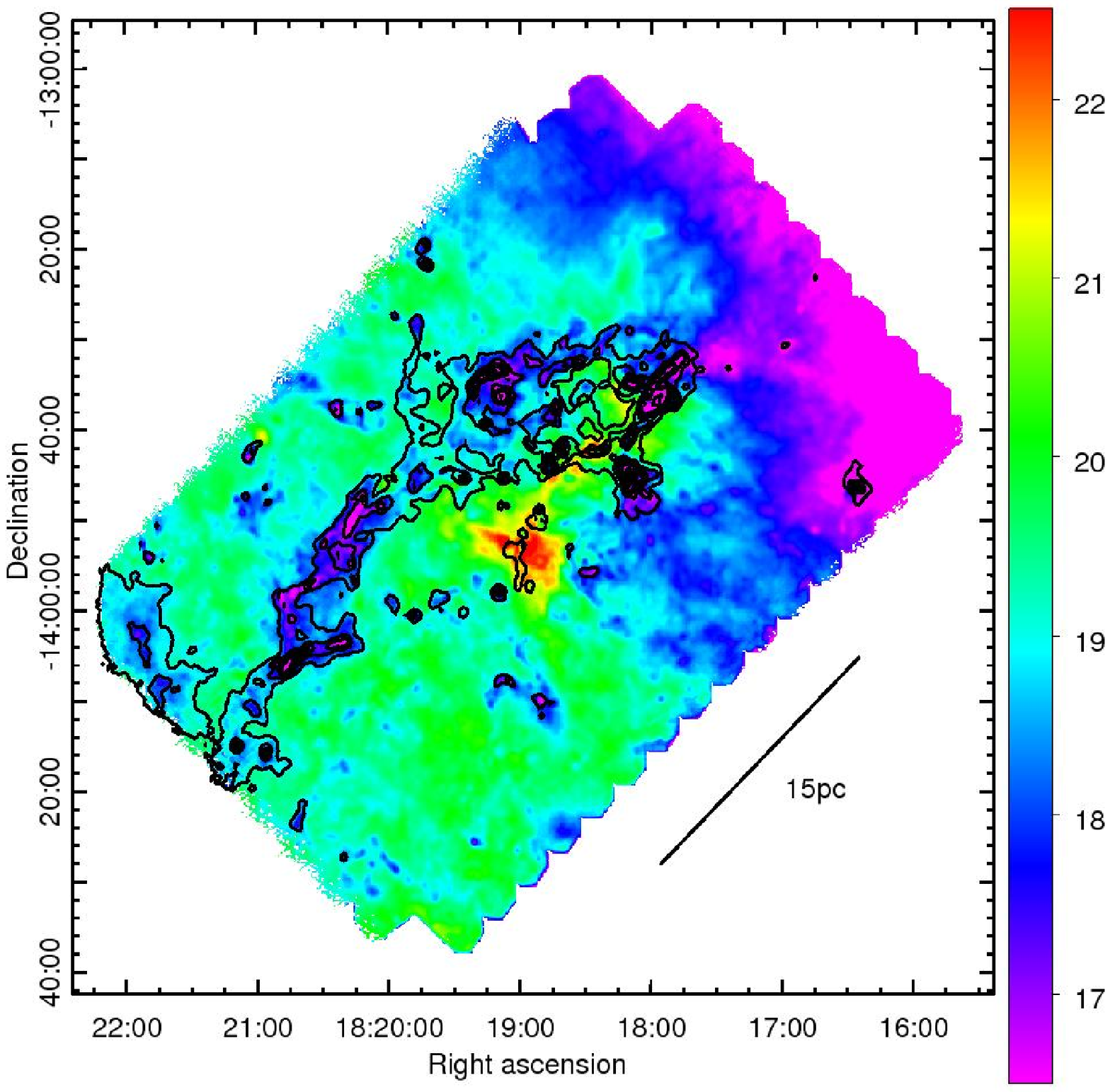}
\caption{Left: Column density map of M16  (37\arcsec\ resolution)
 Right: Dust temperature map, with the column density contours (left) overlaid (37\arcsec\ resolution). The contours are 2.0, 3.1, 4.3, 5.4, 6.5\,$\times$\,10$^{22}$, 1.1, 1.6, 2.0, 2.5\,$\times$\,10$^{23}$ \cmsq. \label{fig:colden}}
\end{figure*}

\section{Observations and data reduction}\label{sec:obs} 

M16 was observed on 2010 March 24 and 2010 September 11\,--\,12, as part of the HOBYS key program \citep{motte10}. In March, the SPIRE \citep[250\,\um, 350\,\um, 500\,\um;][]{griffin10} camera was inadvertently turned off and thus only the shorter PACS \citep[70, 160\,\microm;][]{pog10} wavelengths were mapped during this period. In September, the parallel-scan mode of \herschel\ was used to map simultaneously with both instruments at the aforementioned five bands, using the slow scan speed (20\arcsec/s).
For these September observations, the SPIRE camera was aligned with the PACS data taken from March, allowing 100\% common area coverage (normally, the positional offset between the PACS and SPIRE instruments results in a small (20\arcmin) spatial offset between the PACS and SPIRE maps), as well as an extended PACS coverage to the east and south (see Fig. \ref{fig:hipe:all}, top). A total area of $\sim$~1.5 deg by 1.5 deg was mapped using two orthogonal scan directions. 

Each of the March and September data were reduced with version 5.0.1975 of the \herschel\ Interactive Processing Environment \citep[HIPE\footnote{http://herschel.esac.esa.int/HIPE\_download.shtml};][]{ott10} adopting standard steps of the default pipeline to Level 1 products including calibration and deglitching.
To improve the baseline subtraction, data taken during the turn-around of the telescope were used. Calibration of the PACS data has been found to be within 5\%,  whilst SPIRE calibration is within 10\% for all bands (see the respective observers' manuals).

Maps were produced using the HIPE Level 1 data and v7 of the {\it Scanamorphos} software package\footnote{http://www2.iap.fr/users/roussel/herschel/index.html} which performs baseline and drift removal before regriding \citep{roussel12}. The two sets of PACS data, from March and September, were combined within {\it Scanamorphos} using the `/galactic' option, to produce a single map of the entire PACS coverage. For simplicity, we hereafter 
use PACS data to refer to the combination of the two individual PACS observations, rather than referring to one particular data set.
The individual \herschel\ images at each wavelength can be found in Fig. \ref{fig:hipe:all}, including the
 highest resolution 70\,\um\ image which  traces the hot dust and protostars in this region. The three colour image  of the the entire M16 complex, as well as a zoom to the central region around the OB cluster is given in Fig. \ref{fig:hipe}. 

\section{Structural analysis}\label{sec:structure}

The \herschel\ maps of the Eagle Nebula show bright emission throughout, particularly around the NGC\,6611 cluster (see Fig.~\ref{fig:hipe} and Fig. \ref{fig:hipe:all}). This bright emission covers the Pillars of Creation and the Spire\footnote{The Spire is a well known region in the Eagle Nebula. See, http://en.wikipedia.org/wiki/Messier\_16} east 
of the cluster and a possible ridge (see Section \ref{sec:ring})  of material to the north-west.

Immediately to the north of this ridge is a prominent arch-like structure (see Fig. \ref{fig:hipe}, left). While weak diffuse emission can be seen toward this arch-like structure in the infrared \citep[8\,\um; see Fig.~5 of][]{indebetouw07}, it is much more visually striking and associated with a clear density enhancement in our \herschel\ images e.g., Fig. \ref{fig:hipe}, right (see also Fig. \ref{fig:colden} and Fig. \ref{fig:dt70um} below).
 To the south-west of the ridge is the massive YSO (MYSO) identified by \citet[][also known as IRAS\,18152-1346]{indebetouw07} as the most luminous YSO  (1\,$\times$\,10$^3$\,\lum) in the region. The \herschel\ images of M16 indicate many other sites of star formation, such as small clusters of sources in the south-east of the map, which are more easily identified in the higher resolution PACS data (see Fig. \ref{fig:hipe:all}, top).

The three-colour composite image of this region (Fig. \ref{fig:hipe}) shows a very distinct temperature gradient away from the NGC\,6611 cluster. This temperature gradient is reminiscent of that seen in other HOBYS regions such as  RCW\, 120 \citep{zavagno10} and the bubble \hii\ region study of \citet{anderson12}.

\subsection{Column density \& dust temperature maps}\label{sec:cddt}

\begin{figure}
\includegraphics[height=0.45\textwidth]{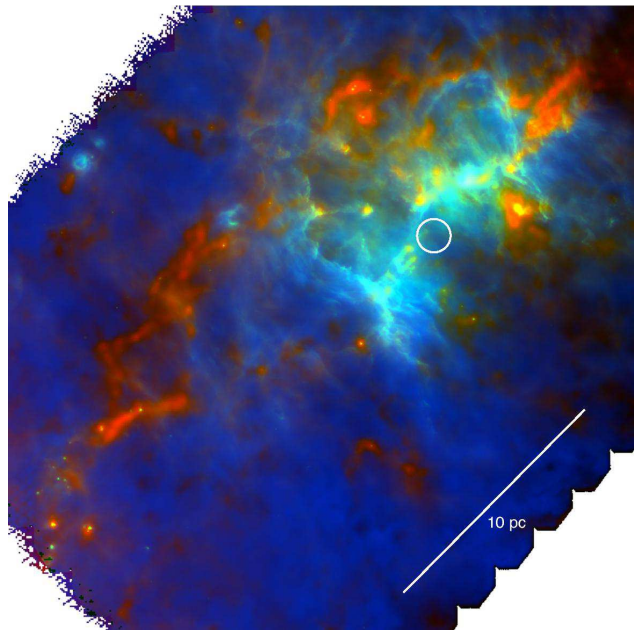}
\caption{Three colour composite of the column density (red), dust temperature (blue) and 70\,\microm\ (green) images. The white circle indicates the location of the NGC\,6611 cluster. The extent of the influence of the photo-dissociation region, created by the presence of the OB cluster, is clearly highlighted in green at the centre of the map. Limb-brightening  effects  occur here around column density peaks and on many low column density filaments around the OB cluster. \label{fig:dt70um}}
\end{figure}

The dust temperature and column density N$_{H_2}$) maps of M16 were drawn by fitting, pixel-by-pixel, spectral energy distributions (SEDs) using a modified blackbody model \citep{hill09, hill10}, following the procedure outlined by \citet{hill11}. 
These data were convolved to the resolution of the 500\,\microm\ band (37\arcsec), and the zero offsets, which were determined from comparison with \Planck\ and \IRAS\ following the procedure detailed by \citet{bernard10}, were applied to the individual maps prior to fitting.
Only the four longest \herschel\ wavebands, which trace cold dense material, were used to make the column density and dust temperature maps (Fig. \ref{fig:colden}), a method consistent within the HOBYS program \citep{hill11, quang11}. The 70\,\um\ emission is excluded from the fits as it is not tracing the cold dust that we are most interested in, and is likely tracing small grains in hot PDRs instead (cf. Fig. \ref{fig:dt70um}).
Furthermore, excluding the 70\,\microm\ emission ensures that the SED fit is more robust, since the 70\,\um\ flux does not completely fit a single temperature greybody model dominated by longer wavelength fluxes.
We used a dust opacity law in which the dust opacity per unit mass column density ($\kappa_\nu$), is given by 
$\kappa_{\nu} = 0.1~(\nu/1000~{\rm GHz})^{\beta}$ cm$^2$/g \citep[e.g.][]{motte10}, and a dust spectral index of $\beta$\,=\,2.
 The latter was fixed in line with the literature  \citep[][and references therein]{goldsmith97} and because we do not have longer wavelength (sub)millimetre data to constrain $\beta$ accurately.
The quality of a SED fit was assessed using $\chi^2$ minimisation. Most of the individual single temperature SED fits comprising the dust temperature and column density maps peaked around, or longward of, the 160\,\um\ flux - the shortest wavelength data used in the fit. 
The observed rms of cirrus noise in the derived column density map is  $\sim$\,2\,$\times$\,10$^{21}$\,\cmsq\ (\Av~$\sim$ 2\,mag). Note however, that the column density and dust temperature are line of sight measures.

The median temperature throughout M16 is $\sim$\,18\,K,
with slightly cooler temperatures in the northern part of the \herschel\ map, which is furthest from the Galactic Plane.
The eastern portion of M16 is dominated by a dense filament (3 -- 7\,$\times$\,10$^{22}$ cm$^{-2}$) situated directly east of, and running parallel to the Pillars of Creation. This `Eastern Filament' appears as a cool temperature depression ($\sim$16\,K; Fig. \ref{fig:colden}, right, see Section \ref{sec:ef}) with respect to the warmer surrounding material.

Figure \ref{fig:dt70um}, a three colour composite of the column density, dust temperature and 70\,\microm\ image, highlights the region affected by strong PDR emission, i.e., the interface between the cloud and the cavity swept out by the NGC\,6611 cluster. The inclusion of the 70\,\um\ emission here highlights clear examples of limb brightening, where the filamentary structures around the cavity have heated edges at the ionisation front created by the NGC\,6611 cluster. Furthermore, this Figure demonstrates the strong correspondence between high column density and low temperature,  especially with respect to the Eastern Filament (clearly visible here in red).

\subsection{Filamentary structure}\label{sec:disperse}

Recent \herschel\ observations have revealed the ubiquity of interstellar filaments, especially in star-forming regions. \citet{andre10, andre11} proposed that in low-mass star-forming regions only those filaments above an \Av\ of $\sim$\,7\,mag are gravitationally unstable (supercritical) and thus capable of forming prestellar cores in their interiors. In fact, a large portion of the molecular complexes observed as part of HOBYS are above this star-forming threshold \citep[e.g.,][]{hill11, quang11, schneider12}.

\begin{figure}
\includegraphics[height=0.5\textwidth,angle=270]{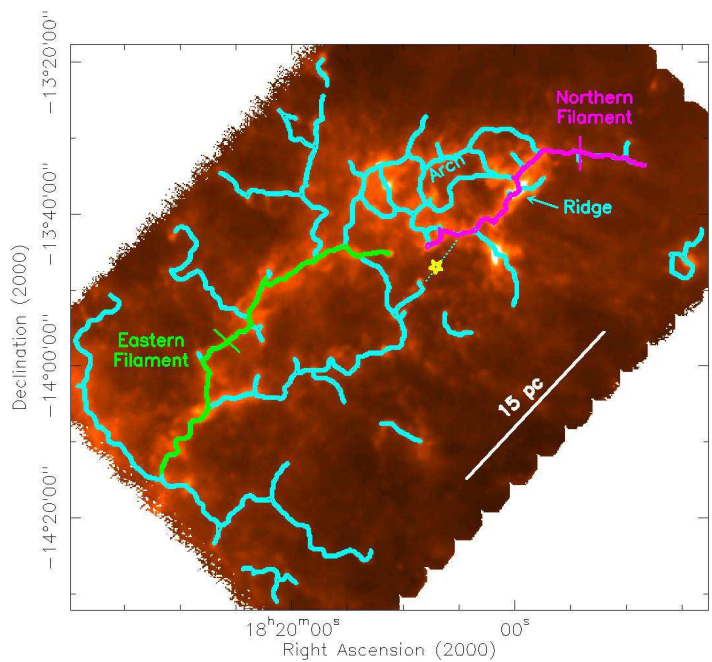}
\caption{Filaments (cyan) in M16 detected by \disperse\ overlaid on the column density map. The Eastern Filament is shown here in green, whilst the Northern Filament is in magenta. The position of the NGC\,6611 cluster is depicted by the yellow star. The temperature minimum of the Eastern Filament and the end of the temperature gradient for the Northern Filament, seen in Fig. \ref{fig:efnf:profile}, are indicated by lines perpendicular to the respective filaments. The possible Western Filament (see Section \ref{sec:disc1}) runs parallel to the Eastern Filament, and the 
possible continuation to the Northern Filament is indicated by a dashed line.   \label{fig:disperse}}
\end{figure}

To take a census of the filaments in the M16 star-forming complex, the \disperse\ algorithm \citep{sousbie11, sousbie11b} was applied to the column density map. \disperse\ works with Morse theory and the concept of persistence to identify topological structures \citep[see][for a detailed explanation of these concepts]{sousbie11} such as filaments and ridges, and connect their saddle-points with maxima by integral lines. This algorithm has already been successfully applied by our group to trace filamentary structures in star-forming regions \citep[e.g., see][]{hill11, peretto12}. The crest points of the filaments, as identified by \disperse\
are presented in Fig. \ref{fig:disperse}. 

The southern half of M16 is dominated by two main filaments which run roughly parallel to each other near the centre of the \herschel\ map, before converging to a solitary filament near the southern edge of the map. The eastern one (`Eastern Filament, see Section \ref{sec:ef}, Fig. \ref{fig:disperse}) is much more prominent (by a factor of $>$\,2 with respect to column density, see also Fig. \ref{fig:colden}, left) than the western one.
The Pillars of Creation lie on a filament\footnote{Note that \disperse\ identifies a single filament for thee pillars, from the column density crest points. This is simply a limitation of the column density map, as at 37\arcsec\ resolution only the densest (western most) pillar is detected (see Fig. \ref{fig:colden}, right).}, which joins the aforementioned Western Filament  (see Fig. \ref{fig:disperse}).

The northern half of the M16 region, just north of the central NGC\,6611 cluster, displays a complex and clustered network of filaments.
The most coherent filament in this region starts close to the NGC\,6611 cluster, connects to the ridge and thereafter heads north-west (see Fig. \ref{fig:disperse}). We hereafter call this filament the `Northern Filament' - see Section \ref{sec:nf}.

Interestingly the clear 70\,\um\ filaments, seen to the south-east  (see Fig. \ref{fig:hipe:all}, top-left), which highlight PDR or low column density filaments, are generally perpendicular to the main SE-NW direction of the cloud and the major filaments (see Fig. \ref{fig:hipe:all}, top-left). These 70\,\um\ emission features are not associated with point like emission of YSOs, but rather to heating/UV ionisation of cloud edges, such as limb brightening/PDRs.
These filaments are similar to the striations seen in regions such as Taurus \citep{pedro12}.

\subsection{The arch and a potential  ridge}\label{sec:ring}

In the individual \herschel\ images of the Eagle Nebula there is clear evidence of a arch-like structure to the north-west of the NGC\,6611 cluster (Fig. \ref{fig:hipe} and Fig. \ref{fig:hipe:all}). This arch-like structure is actually composed of two arches, an inner warmer ($\sim$\,17\,--\,20\,K) arch, and an outer high column density ($\sim$\,2.0\,--\,5.8\,$\times$\,10$^{22}$\,\cmsq)  arch.
 These two arches may in fact be tracing different parts of the same shell-like structure or bubble, as discussed below in Section \ref{sec:disc1}.  These structures are more prominent at the shorter PACS wavelengths. Due to its tendency of tracing density crests, \disperse\ only traces the column density arch (see, Fig. \ref{fig:disperse}).

Since the arch appears weak at 8\,\um\  
\citep[with IRAC; see][Fig. 5]{indebetouw07}, and not as clearly as in our \herschel\ images (see for example Figs. \ref{fig:hipe} and \ref{fig:hipe:all} and Section \ref{sec:structure}), it is likely composed of mostly cool material. 
At the base of the arch is a potential ridge of material. \citet{hill11} define a `ridge' as a high-column density ($\sim$10$^{23}$\cmsq) super-critical filament dominating its environment that is capable of forming high-mass stars. The most prominent part of the Northern Filament  has a  column density above 1.0$\times$10$^{23}$\cmsq\ over $\sim$\,2\,pc. It is not as dominant, i.e., containing most of the material in the vicinity,  as the Vela~C ridge or DR\,21 ridge \citep[][Hennemann et al., in prep, respectively]{hill11}.
This ridge and its star formation content is discussed by \citet{white12}.

\section{Impact of the NGC\,6611 cluster on the M16 cloud}\label{sec:discuss1}

The NGC\,6611 cluster is responsible for most of the mechanical and radiative
energy incident on the molecular cloud complex around the Eagle Nebula (M16).
Here we assess the dust heating effects in view of constraining the geometry
of the complex.

\begin{figure*}
\includegraphics[height=7cm, width=9cm]{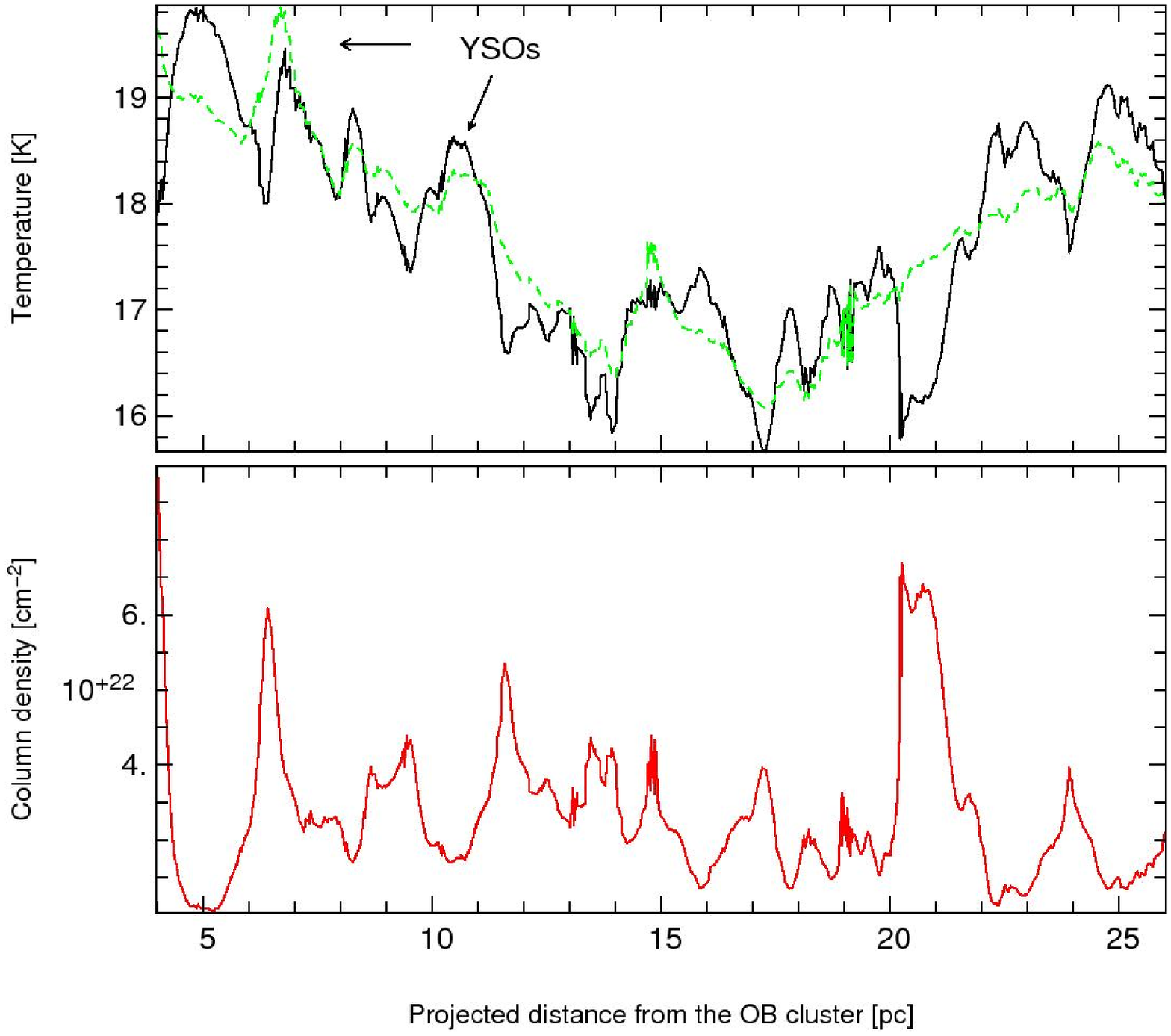}
\hfill
\includegraphics[height=7cm, width=9cm]{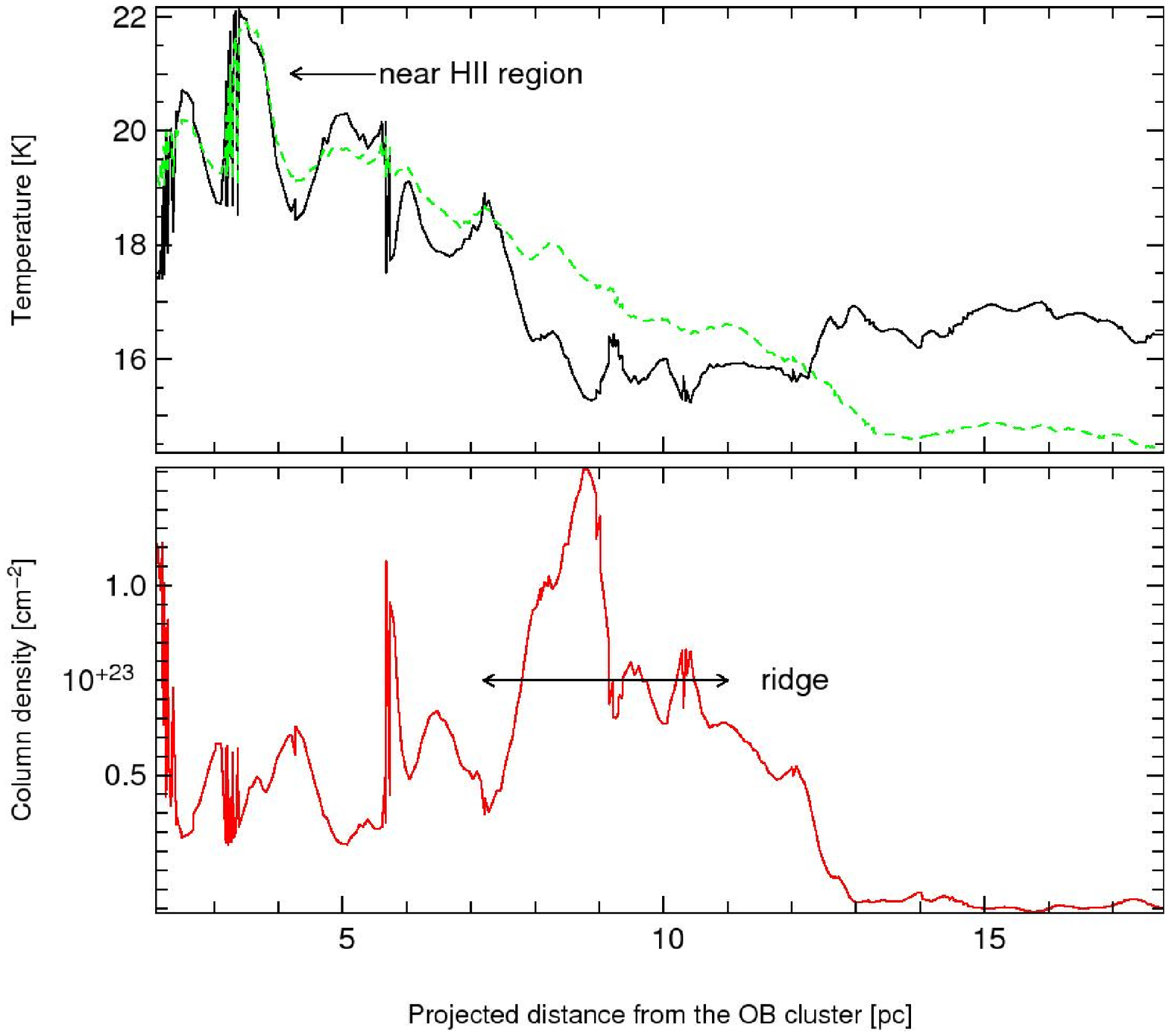}
\caption{The dust temperature (top) and column density (bottom) profiles along the Eastern (left) and Northern (right) Filaments. The positions along the filaments are plotted as a function of their projected distance from the cluster. The NGC\,6611 cluster is located at `0' on the x-axis, centred at 18 18.8 -13 49. The green dashed line on the dust temperature (top) is the dust temperature after removing the effects of cold sources within the filaments, following the relation of \citet[Fig. 1;][see footnote 6 and also Didelon et al., in prep. who examine this relation for \herschel\ filaments]{juvela11b}. The thick vertical lines seen on the profile of the Northern Filament indicates structures at the same distance from the cluster, as the filament tends to curve around the cavity.
\label{fig:efnf:profile}}
\end{figure*}

\subsection{The cold Eastern Filament}\label{sec:ef}

The Eastern Filament is a cold prominent filament running east of the NGC\,6611 cluster in the column density and dust temperature maps (Figs. \ref{fig:colden}, \ref{fig:dt70um} and \ref{fig:disperse}).
This southern portion of this filament houses many 70\,\um\ sources (see Fig. \ref{fig:hipe:all}, top-left), that we interpret as embedded protostars. 
In order to determine how the presence of the OB cluster in M16, in particular its heating  effect, has impacted the Eastern Filament, column density and dust temperature profiles were taken along the filament. These profiles were derived using the crest points identified by the \disperse\ algorithm (see Section \ref{sec:disperse}), and are presented in Fig. \ref{fig:efnf:profile} (left) as a function of projected distance from the cluster.
The dust temperature profile of the Eastern Filament shows a  decreasing temperature with increasing distance from the OB cluster. There are small fluctuations along the temperature profile, which are anti-correlated with column density (see the lower panel of Fig. \ref{fig:efnf:profile}) and thus density enhancements such as star-forming sources located within the filament.
Around $\sim$\,17\,pc from the OB cluster the temperature of the Eastern Filament stabilises as it approaches the southern edge of the map.
At distances $>$\,17\,pc, the dust temperature in the Eastern Filament increases with increasing distance as it approaches the Galactic Plane. 
The Eastern Filament has a $\sim$\,3\,K temperature gradient over its projected length of $\sim$\,20\,pc. 

To remove the contributions of cold dense sources within the Eastern Filament to the dust temperature we have a applied a de-screening to the data following the relation\footnote{T$_{corrected}$~=~T$_{measured}$ + 4.4 log (N$_{H_2}$/
N$_{H_2}$\,${\rm _{ref}}$); where N$_{H_2}$\,${\rm _{ref}}$ is the median column density of the filament. This relation, found in radiative transfer simulations of dense cores, is taken here as an approximation to that which should exist in Plummer-like \citep[cf.][]{arzoum11} filaments. A paper is currently being prepared on this topic which examines such a relation in \herschel\ filaments \citep{didelon12}.}
 by \citet[Fig. 1;][]{juvela11}. This relation (green line on Fig. \ref{fig:efnf:profile}) simulates the temperature structure for a filament with a homogeneous density crest, and  
 and  while it removes the effects of density enhancements above N$_{H_2}$\,${\rm_{ref}}$ (where  N$_{H_2}$\,${\rm _{ref}}$ is taken here as 3.2\,$\times$\,10$^{22}$\,\cmsq) including, cold dense fragments embedded within the filament, it does not remove the centrally heated protostars (indicated as YSOs on Fig. \ref{fig:efnf:profile}).  This column density filtering highlights the effect of the OB cluster on the dust temperature of the Eastern Filament as a whole, the overall trend of which is a decreasing dust temperature with increasing distance from the OB cluster up to a projected distance of $\sim$\,17\,pc.

In contrast to the dust temperature, the mean column density profile (Fig.~\ref{fig:efnf:profile}, left) appears to remain  rather constant with  distance from the OB cluster, though it is subject to the aforementioned density enhancements. The strong (and wide) density fluctuation seen at $\sim$\,20\,pc reaching 7\,$\times$\,10$^{22}$\,\cmsq, is located at approximately where the Eastern and Western Filaments 
 appear to meet (see Section \ref{sec:disperse}). 

\begin{table}
\caption{The extent of the heating penetrating of NGC\,6611 on the two main filaments identified in the M16 complex, determined from the dust temperature profiles (Fig.~\ref{fig:efnf:profile}).
The parameters of the column density and dust temperature derived from probability distribution functions (PDFs) for each sub-region in M16 (Section \ref{sec:pdf} and Fig. \ref{fig:pdf})  as well as their heating distance and gradient are also given (Fig. \ref{fig:rp:all}).
The numbers given in parentheses in the last column is the value obtained when a de-screening has been applied (see section \ref{sec:ef}). The asterisk here ($^\ast$) indicates that the heating gradient was taken over the steepest portion (3\,--\,5\,pc), see Fig. \ref{fig:rp:all}. \label{tab:4band:pdf}}
\begin{tabular}{@{}l@{}c@{}c@{}cc@{}c@{}c}
\hline
Name  & Colour & Column & Temper- & Penetration & {Heating}\\
& & Density & ature& Distance & {Gradient}\\
& &  ($\times$\,10$^{22}$\,\cmsq) & ~(K) &  (pc) & (K/pc)\\
& &median & median & (heating) & (filaments) \\
\hline
\hline
E. Filament & green &  &   & 15 & 0.25 (0.23)\\  N. Filament & magenta & & & 10 & 0.63 (0.45)\\ \hline
M16-N & green &  2.3 &  18.7 & 9 & 0.38 \\
P. of Creation & blue &  1.4 &  20.3 & 3 & flat\\
Spire & cyan & 1.9 &  19.7 & 6 & 0.24\\
MYSO & magenta & 2.0 &   18.0 & 5 & 2.25$^\ast$\\
M16-E & red & 1.2 & 18.9 & -- & --\\
M16-NW  & yellow & 1.2 & 17.2 & 9 & --\\
\hline
\end{tabular}
\end{table}

\subsection{The Northern Filament}\label{sec:nf}

The Northern Filament originates at the eastern part of the NGC\,6611 cavity, runs around the cavity and across the potential ridge \citep[see Section \ref{sec:ring} and][]{white12} before continuing into the northern part of the map (see Fig. \ref{fig:disperse}). 
The column density and dust temperature profiles of the Northern Filament are given in Fig. \ref{fig:efnf:profile} (right). The dust temperature of this filament remains rather constant ($\sim$20\,K) around the NGC\,6611 cluster, except for a temperature spike at $\sim$\,3\,pc corresponding to a position close to a well known \hii\ region (IRAS 18156-1343). At $\sim$\,6\,pc the dust temperature drops rapidly with increasing distance from the NGC\,6611 cluster (5\,K over 4\,pc), before stabilising at a near constant temperature of $\sim$\,16\,K. This temperature drop is greater than that seen in the Eastern Filament (see Fig. \ref{fig:efnf:profile}, left). This difference is (partly)  because the column density profile is less homogeneous than the Eastern Filament, with a large dip in the column density curve due to the presence of the ridge. While the dust temperature used here is a line of sight average for the filament, and the absolute values of the dust temperature are likely to be slightly overestimated as a result, the gradient identified in these Eastern and Northern filaments is still present and robust.

Similarly to the Eastern Filament, a de-screening was applied to the Northern Filament, with N$_{H_2}$\,${\rm _{ref}}$, the median column density, taken as 4.4\,$\times$\,10$^{22}$\,\cmsq. After applying this de-screening,  the dust temperature of the Northern Filament (green line Fig. \ref{fig:efnf:profile}, right) decreases steadily over a much greater distance (5\,K over 8\,pc). 
 Even after de-screening, the drop in dust temperature is greater than that in the Eastern Filament, which 
may result from the less homogeneous column density in the Northern Filament or
it might also suggest a larger relative inclination along the line of sight, though spectral line (velocity) information is needed to support this idea.
Note that at distances $>$\,12\,pc the de-screening temperature is much lower ($\sim$\,2\,K) than that without de-screening. This is likely due to the fact that there are very few dense cores in this portion of the filament (see Fig. \ref{fig:colden}), and thus the de-screening does not  completely apply here.

\begin{figure*}[]
\includegraphics[height=0.285\textwidth]{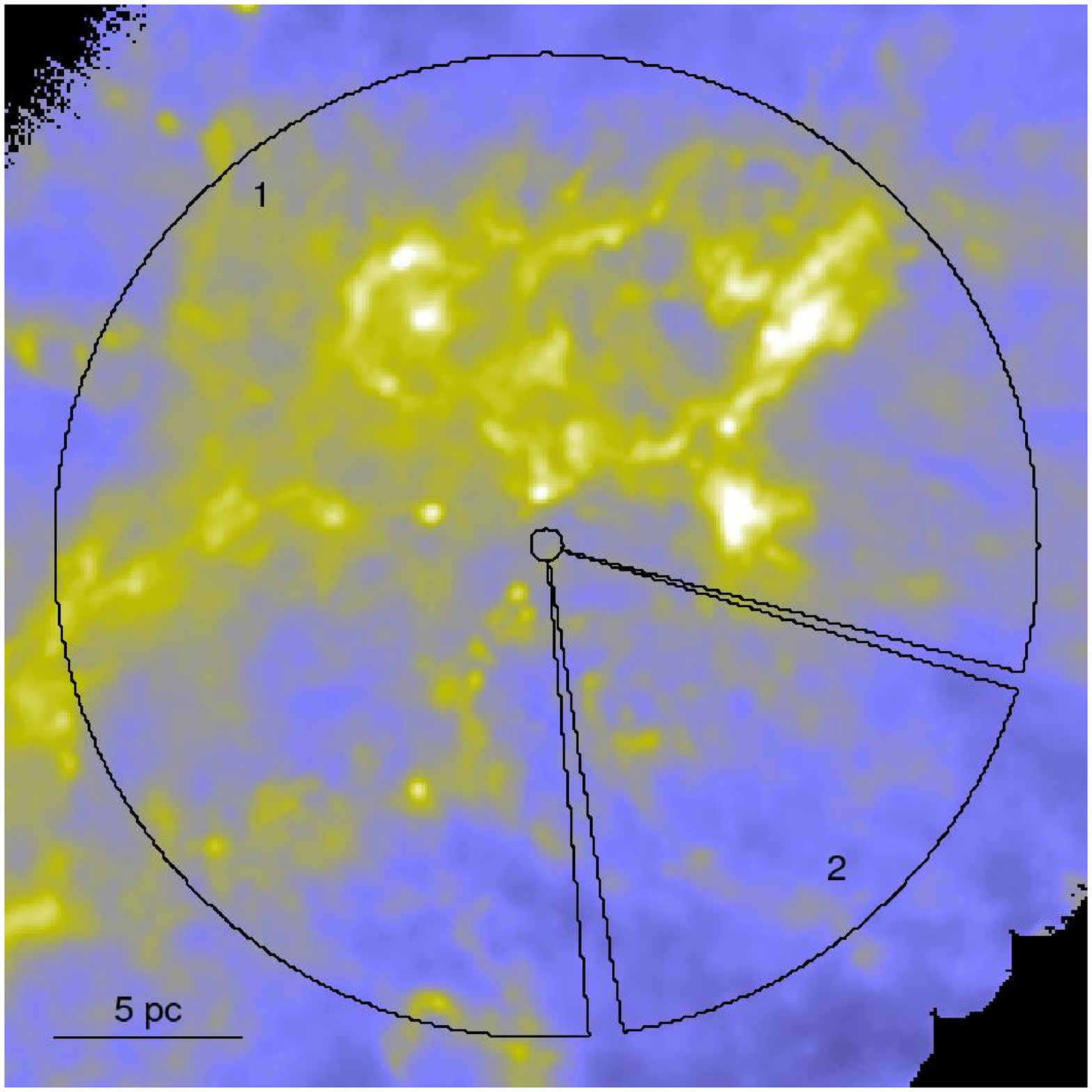}
\hfill
\includegraphics[height=0.285\textwidth]{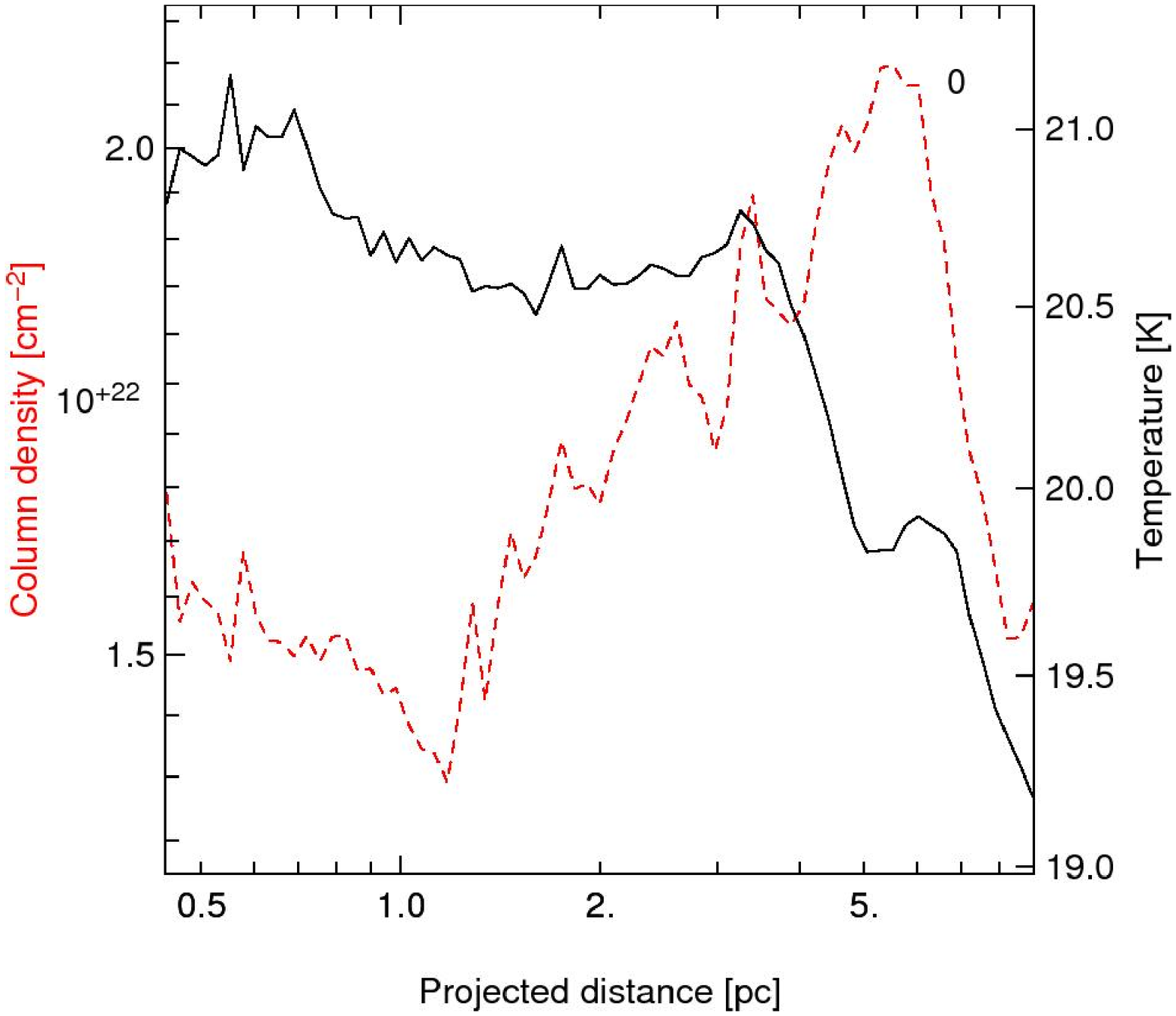}
\hfill
\includegraphics[height=0.285\textwidth]{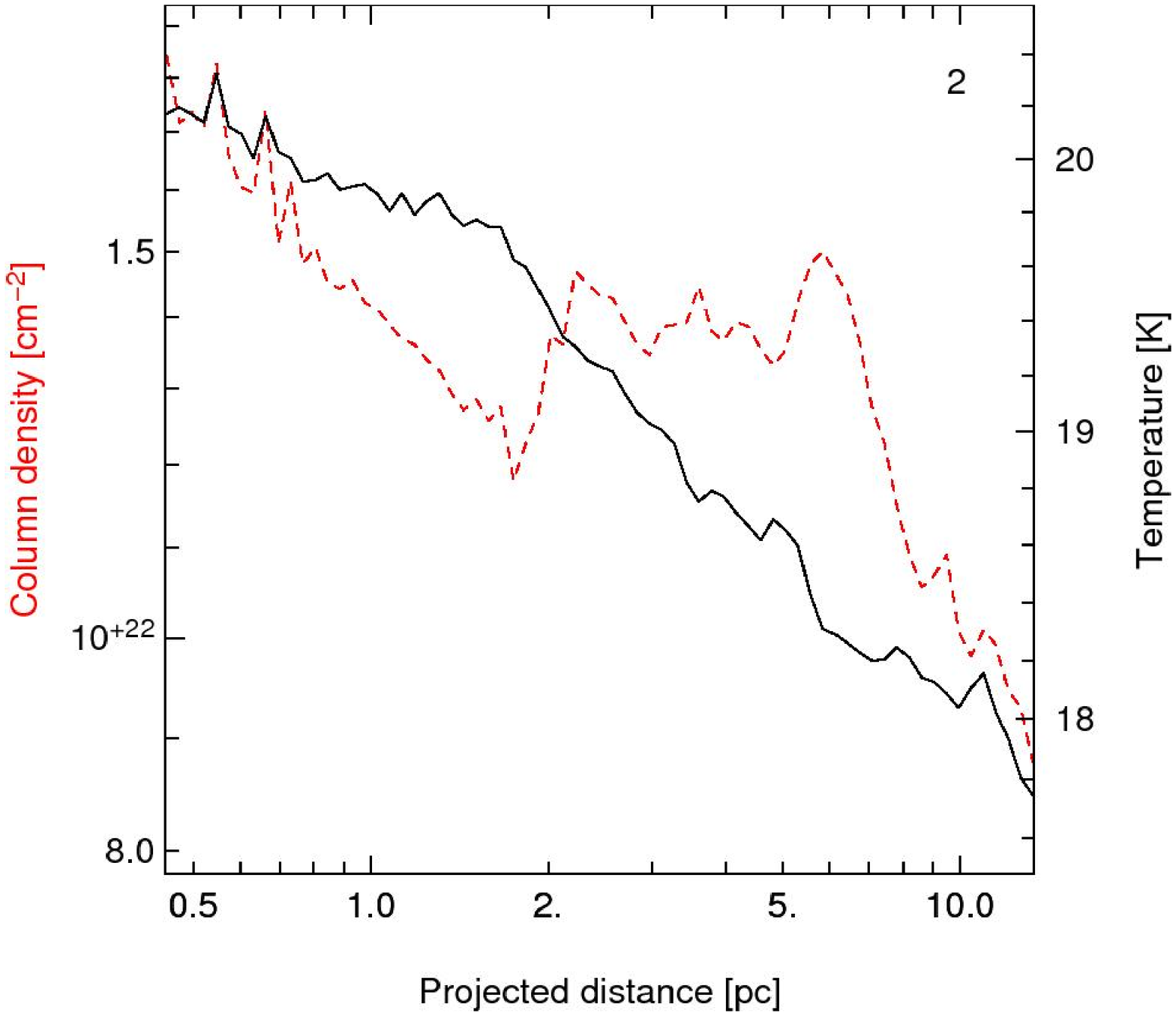}
\caption{Left: Column density map of NGC\,6611 with radial segments as indicated. Segment 1 (middle) contains all of the star formation activity in the region, and segment 2 (right) the cavity carved out by the NGC\,6611 cluster. These profiles have been averaged over azimuthal angle. The black solid line is the dust temperature profile, while the red dashed line is the column density profile. More detailed segments for each region inside segment 1 are given in Fig. \ref{fig:rp:all}. Cuts are the median profile in each segment to limit biases introduced by the presence of protostars. \label{fig:rp}}
\end{figure*}

\subsection{NGC\,6611 and its immediate surroundings}

In order to determine the extent of the heating from the OB cluster on low column density material (1.5\,$\times$\,10$^{22}$\,\cmsq\ or lower, see Fig. \ref{fig:rp}), PA-averaged radial profiles of both the dust temperature and the column density were taken covering all of the dense star-forming material in the region, as well as the cavity carved out by the NGC\,6611 cluster where there is more diffuse material. 
These two radial profiles are presented in Fig. \ref{fig:rp}, with narrower radial profiles covering the interesting features in this region, e.g., the pillars, given in Fig. \ref{fig:rp:all}. 
The profiles, which take the median profile in the region rather than the average to limit the inclusion and bias of protostars, were centred at the position of the NGC\,6611 cluster. 
It is immediately clear (see Fig. \ref{fig:rp}) that the dust temperature is rather homogeneous close to the NGC\,6611 cluster ($\sim$\,1.5\,--\,2\,pc) but then it drops with increasing distance from the cluster. The dust temperature profile covering most of the emission in the region (Segment 1, Fig. \ref{fig:rp}) drops 2\,K within $\sim$\,4\,pc, and continues to decrease with increasing distance. The dust temperature profile of the cavity (Segment 2) indicates a homogeneous dust temperature close to the cluster ($<$\,2\,pc), and then a steadily decreasing temperature with increasing distance from the cluster.
The column density is anti-correlated with the dust temperature, a result which also applies to the smaller segments presented in Fig. \ref{fig:rp:all}. Each of the segments presented in Fig. \ref{fig:rp:all} display different dust temperature profiles, suggesting different column densities of material and/or three-dimensional effects.

The extent of heating by the NGC\,6611 cluster on the surrounding environment, i.e., the penetration distance, is found to be 3\,pc to 15\,pc as determined from the dust temperature profiles (see the dust temperature gradient in Figs. \ref{fig:efnf:profile}, \ref{fig:rp}, \ref{fig:rp:all}), depending on the region (see Table~\ref{tab:4band:pdf}).
It is not qualitatively different from the penetration measured for 3\,--\,6\,$\times$\,10$^{22}$\,\cmsq\ filaments (see Section \ref{sec:ef} and \ref{sec:nf}).

\section{Discussion \& Interpretation}

The Eagle Nebula (M16) is arguably one of the finest examples of high-mass star formation in the Galaxy, chiefly because of the large number of high-mass stars forming in this region \citep[cf.][]{hillenbrand93} and the remarkable visual effects that this activity has had on the complex (e.g. the Pillars of Creation, Spire, and the `Eagle' in the optical). The Nebula's centre has largely been excavated by the radiation and powerful stellar winds of the OB stars within the central NGC\,6611 cluster.
This radiation has shaped, and continues to shape, the molecular cloud material, possibly even triggering the formation of new stars and the dissipation of material.

\subsection{The (three-dimensional) structure of the complex}\label{sec:disc1}

The three-colour composite image of this region (Fig. \ref{fig:hipe})  shows a very distinct temperature gradient running away from 
the centre of the cavity carved out by the NGC\,6611 cluster. 
Those regions closest to the cluster are subject to heating by the OB stars, which is reflected in the higher dust temperatures (+3\,--5\,K) compared with those regions in the more remote (14\,--17\,pc) parts of the cloud, where the heating no longer penetrates.
The temperature gradient from the NGC\,6611 cluster is similar to that seen in Rosette \citep[][]{schneider10} but over a smaller distance, with a clear drop in temperature within 10\,pc of the cluster in all directions (see Figs. \ref{fig:efnf:profile} and \ref{fig:rp}).

The M16 complex displays evidence of a large number of bright rims, which typically occur close to PDRs, especially in the vicinity of the NGC\,6611 cluster (see Figs. \ref{fig:colden}, right and \ref{fig:dt70um} and Section \ref{sec:disperse})
Whilst a greater investigation into these potential sites of limb-brightening is warranted, we can use the information in these maps to infer the structure relative to the OB cluster (e.g., Minier et al., in prep). Those objects that exhibit signs of bright blue limb-brightening emission at their surface are likely directly receiving incident radiation from the NGC\,6611 cluster and are thus at the same distance as the cluster or behind it. 
There are clear pockets of hot dust (traced by 70\,\um\ emission, shown in blue, in Fig. \ref{fig:hipe}) on the southern side of the NGC\,6611 cluster, around the pillars, and to the north of the cluster, as well as inside the arch structure (also seen in Fig.~\ref{fig:dt70um}).
This inner 10\,pc shell is suggestive of a three-dimensional geometry where the bright 70\,\um\ objects, such as the Pillars of Creation, are illuminated by the NGC\,6611 cluster and not extinguished by any cloud in its foreground. The location of the Spire and the ridge is less clear.
On the other hand, objects such as the MYSO, which do not show evidence of bright rims, and thus incident radiation from the cluster, are clearly in front of the cluster, and closer to us along the line of sight. 
An interesting question is raised with respect to the origin of the arch structure (Section \ref{sec:ring} and Fig. \ref{fig:disperse}). Is it actually one structure e.g., created by the expansion of radiation similar to that of a bubble, or is it a consequence of morphology? 
 As noted in Section \ref{sec:ring}, two arches are in fact detected at this location and \disperse\ only detects the denser of these. Figure~\ref{fig:hipe} and Fig. \ref{fig:colden} (right) suggest that the inner warmer arch is closer to the cluster, directly influenced by its heating, whilst the outer higher column density arch is likely above  (i.e., closer to us along the line of sight) and away from the cluster. This suggests some sort of curved morphology, where the two arches are in fact two sides of a three dimensional shell-like or bubble-like structure, with the near (denser arch) and far (warmer arch) curved sides of the shell/bubble appearing as two arches in two dimensions.

Two prominent filaments are detected in the M16 complex, a cool dense Eastern Filament (see Section \ref{sec:ef}) which spans almost 25\,pc, and the Northern Filament (see Section \ref{sec:nf}) which extends across the ridge, at the western side of the arch structure, and up to $\sim$\,15\,pc.
Figs. \ref{fig:colden} (left) and \ref{fig:dt70um} indicate a clear branch towards the southern end of the Eastern Filament, at $\sim$\,15\,pc from the origin (see Fig. \ref{fig:efnf:profile}, left), as discussed in Section \ref{sec:ef}. \disperse\ (Section \ref{sec:disperse}) actually detects the much weaker (with respect to column density - Fig. \ref{fig:colden}) western component of this branch (Fig. \ref{fig:disperse}), which extends up to the base of the Pillars of Creation. If we were to extrapolate or extend this western filament, it would continue to join the Northern Filament at the approximate location of, and along the same line of sight as, the MYSO.
In this scenario, the Eastern Filament has remained largely undisturbed whilst the Western Filament has undergone a dramatic transformation. Once the home of the NGC\,6611 cluster, it would have been destroyed as the cluster accumulated mass and formed the high-mass OB stars seen in existence today. The strong ionising winds from the OB stars themselves would have carved out a cavity in this filament, and eradicated any existence of the continuation of this Western Filament. Small grains associated with this Western Filament are still present and emit strongly at 70\,\um\ (Fig. \ref{fig:hipe:all}, top-left).

\subsection{The heating effect of the NGC\,6611 cluster}

The high-mass star cluster NGC\,6611 has noticeably impacted, in terms of heating, the M16 complex as a whole. The dust temperature profiles (Figs. \ref{fig:efnf:profile} and \ref{fig:rp}) clearly show the \emph{effect} and the \emph{extent} of this impact, i.e., how much and how far it has affected the temperature  in all directions. 
Despite the fact that the dust temperature measured at the crest (Fig. \ref{fig:efnf:profile}) or averaged within segments (Fig. \ref{fig:rp}) obviously depends on the column density of the material traced, the  penetration distances do not seem very different from the diffuse emission of segment 1 ($<$\,N$_{H_2}\,>$\,10$^{22}$\,\cmsq) to the high column density emission of the Northern Filament ($<$\,N$_{H_2}~\gtrsim$~6\,$\times 10^{22}$\,\cmsq). 
Our analysis (see Figs.~\ref{fig:efnf:profile} and \ref{fig:rp} and Table\,\ref{tab:4band:pdf}) has shown that NGC\,6611 is able to heat the M16 cloud to a projected distance of $\sim$$8-10$\,pc. The M16 cloud displays $\sim$$1.5-5$~K temperature variations, corresponding to heating gradients of 0.25--0.6\,K/pc, with an average value of $\sim$0.3\,K/pc. 
The dust temperature profiles of the Eastern and Northern Filaments (Fig. \ref{fig:efnf:profile}) indicate that NGC\,6611 is able to heat, by 3\,--\,5~K, star-forming sites ($>$\,10$^5$\,\cmc\ dense cores) approximately up to 10\,--\,11\,pc into each filament. While the dust temperature and column density gradients are robust, the absolute values of these quantities are subject to line of sight effects.

It is clear from the radial profiles (Fig. \ref{fig:rp}) that the heating effect of NGC\,6611 is not the same in all directions. These differences would be expected if the NGC\,6611 cluster is in fact at the same 
distance as only
 some of these regions (as suggested in Section  \ref{sec:disc1}). Alternatively, the ionisation and heating depths of the cluster is simply dependent on the amount of material in the way, such that high column density regions will block the radiation  better than low column density regions in which the radiation more easily penetrates.

\herschel\ observations have shown that interstellar filaments act as the birthplace of star formation.  In M16, we have shown that interstellar filaments can largely be influenced by heating, such as that coming from the nearby  high-mass NGC\,6611 cluster (e.g., Fig. \ref{fig:efnf:profile}).  Such external heating is often not considered when deriving the evolutionary status of a core from its far-infrared SED. Evolution/age proxies such as 
L$_{submm}$/$L_{bol}$  \citep{andre00} 
or T$_{bol}$ \citep{myers93} for evolutionary diagrams have been recently extrapolated to massive dense cores \citep[e.g.][]{motte01}. These parameters should be taken with caution, since massive dense cores generally form in molecular cloud complexes associated with OB clusters. 
By ignoring this effect, T$_{bol}$ will appear cooler at larger distances from the cluster, and hotter closer to the cluster. Similarly, the L$_{submm}$/$L_{bol}$ ratio of a protostellar core will be larger further from the cluster than closer to it.
Inferring evolutionary trends from these parameters, one would then naturally find more evolved protostellar dense cores closer to the cluster than further away from it, giving the (incorrect) impression of triggered star formation.  In order to avoid these biases, the effects of external heating on interstellar filaments should be considered when determining the evolutionary state of a star-forming core, especially in high-mass star-forming regions.

As an example, a $\sim$\,6\,K dust temperature gradient has been found for massive dense cores forming in the Rosette molecular cloud under the influence of the NGC\,2244 cluster \citep{motte10}. Even after subtracting a possible 3\,K gradient coming from the different temperatures of the cloud filaments inside which prestellar cores have formed and evolved into protostars, the remaining 3\,K could still argue for an  age trend for these YSOs, as originally suggested by \citet{schneider10}.

Our \herschel\ data of M16 do not provide any evidence for a supernova remnant, as suggested by \citet{flagey11}. Our maps (Fig. \ref{fig:colden}, right) indicate dust temperatures in M16 that are consistent with other high-mass star-forming regions \citep{quang11, hill11}.  \citet{flagey11}, with their shorter wavelength \spitzer\ data, were more sensitive to hot dust which is being shattered by a supernova remnant. Based on this comparison alone, we are unable to  support or refute this claim based on our \herschel\ data, which are more sensitive to colder dust.
However,  young supernova remnants such as SN\,1987A \citep{matsuura11} and Cassiopeia A \citep{barlow10}  have been readily observed by \herschel, as have much older supernova remnants e.g. W44. SN\,1987A is seen as a point source with PACS and as extended emission at increasingly longer \herschel\ wavelengths, while Cas A is well traced at all \herschel\ wavelengths, as well as at 24\,\um\ with \spitzer.
The fact that we don't see evidence of such a supernova remnant in our individual \herschel\ images, and that we see no evidence of impact on the column density distribution around the NGC\,6611 cluster, as well as the 
absence of emission seen with \spitzer\ \citep[e.g.][]{flagey11, indebetouw07} does not favour the presence of a supernova remnant in this region.

\begin{acknowledgements}
T.H. is supported by a CEA/Marie-Curie Eurotalents Fellowship. D.E., D.P., K.L.J.R., and E.S. are funded by an ASI fellowship under contract numbers I/005/11/0 and I/038/08/0. The authors wish to thank Daniel Price for comments and suggestions on an earlier version of the manuscript. We'd also like to thank Christophe Carreau (ESA) for working his magic with Fig. \ref{fig:hipe}.
 Part of this work was supported by the ANR (\emph{Agence Nationale pour la Recherche}) project `PROBeS', number ANR-08-BLAN-0241. This work has made use of the Yorick freeware package (see http://yorick.sourceforge.net/). Fig. \ref{fig:hipe:all} was generated using APLpy, an open-source plotting package for Python hosted at http://aplpy.github.com. 

SPIRE has been developed by a consortium of institutes led
by Cardiff Univ. (UK) and including: Univ. Lethbridge (Canada);
NAOC (China); CEA, LAM (France); IFSI, Univ. Padua (Italy);
IAC (Spain); Stockholm Observatory (Sweden); Imperial College
London, RAL, UCL-MSSL, UKATC, Univ. Sussex (UK); and Caltech,
JPL, NHSC, Univ. Colorado (USA). This development has been
supported by national funding agencies: CSA (Canada); NAOC
(China); CEA, CNES, CNRS (France); ASI (Italy); MCINN (Spain);
SNSB (Sweden); STFC, UKSA (UK); and NASA (USA). 

PACS has been developed by a consortium of institutes led by MPE (Germany) and including UVIE (Austria); KU Leuven, CSL, IMEC (Belgium); CEA, LAM (France); MPIA (Germany); INAF-IFSI/OAA/OAP/OAT, LENS, SISSA (Italy); IAC (Spain). This development has been supported by the funding agencies BMVIT (Austria), ESA-PRODEX (Belgium), CEA/CNES (France), DLR (Germany), ASI/INAF (Italy), and CICYT/MCYT (Spain).

\end{acknowledgements}

\bibliographystyle{aa} \bibliography{/Users/thill/pap_write/bib/references} 

\begin{appendix} 
\section{\herschel\ images of M16}

\begin{figure*}
\includegraphics[width=0.5\textwidth]{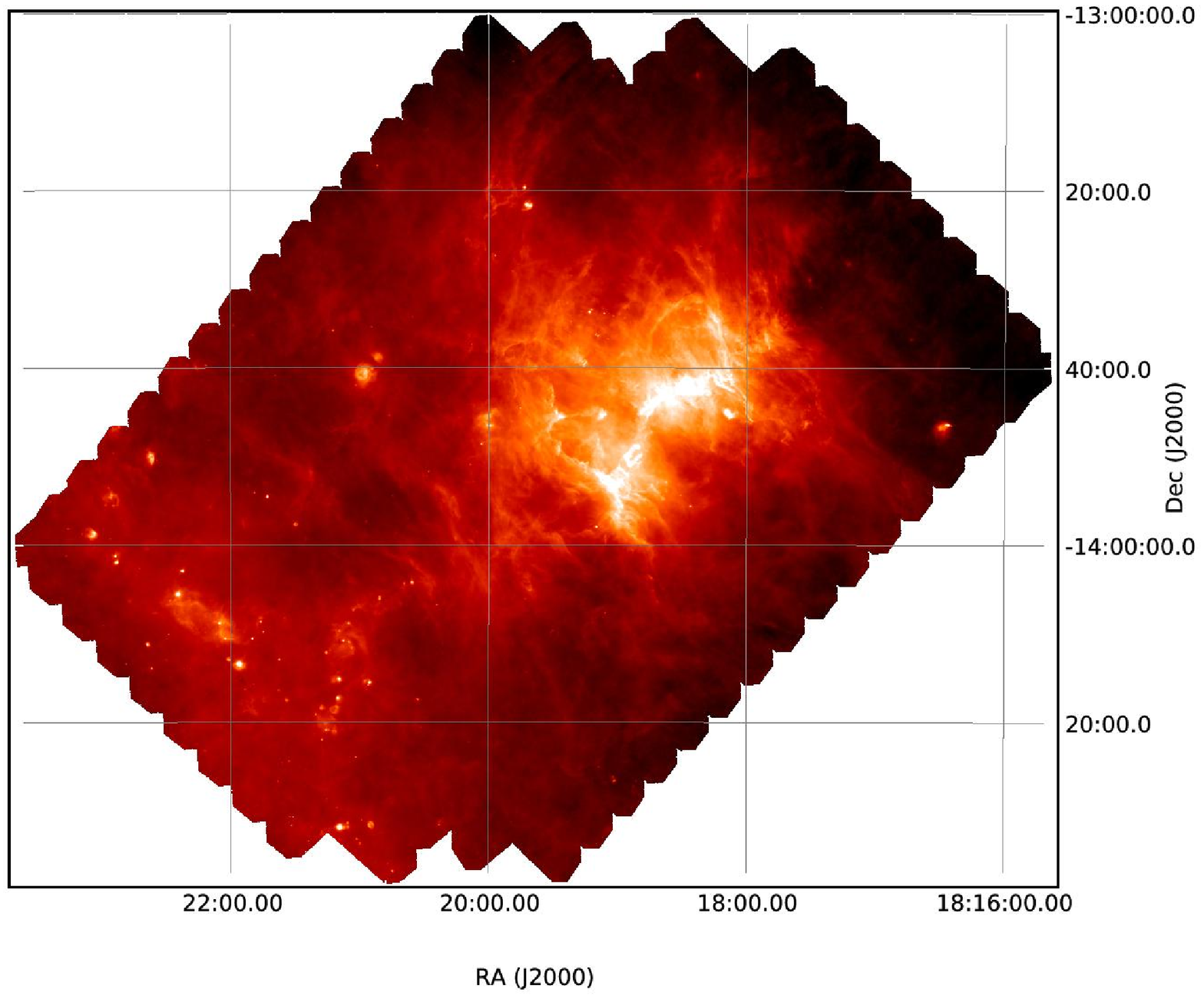}
\includegraphics[width=0.5\textwidth]{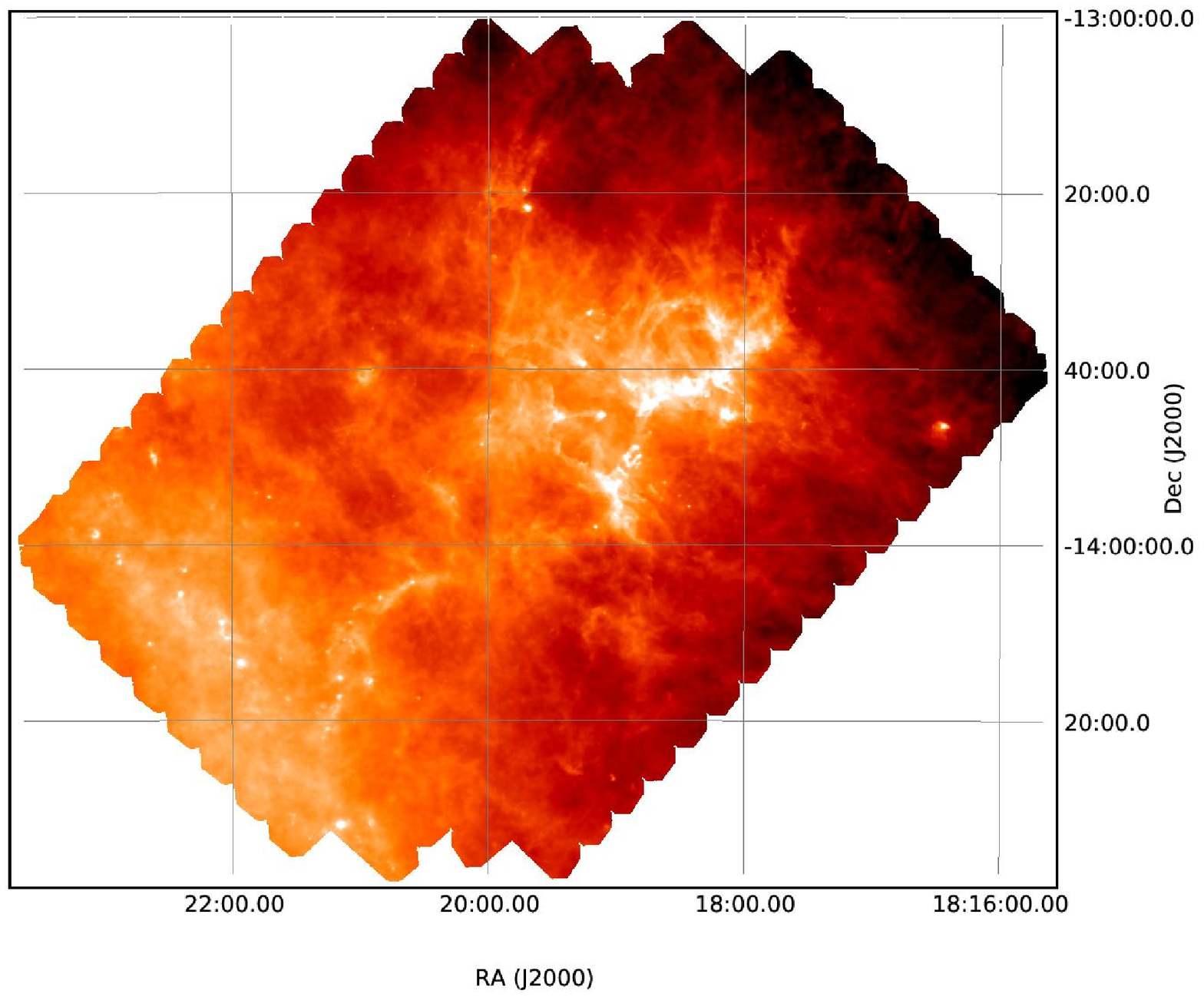}
\includegraphics[width=0.5\textwidth]{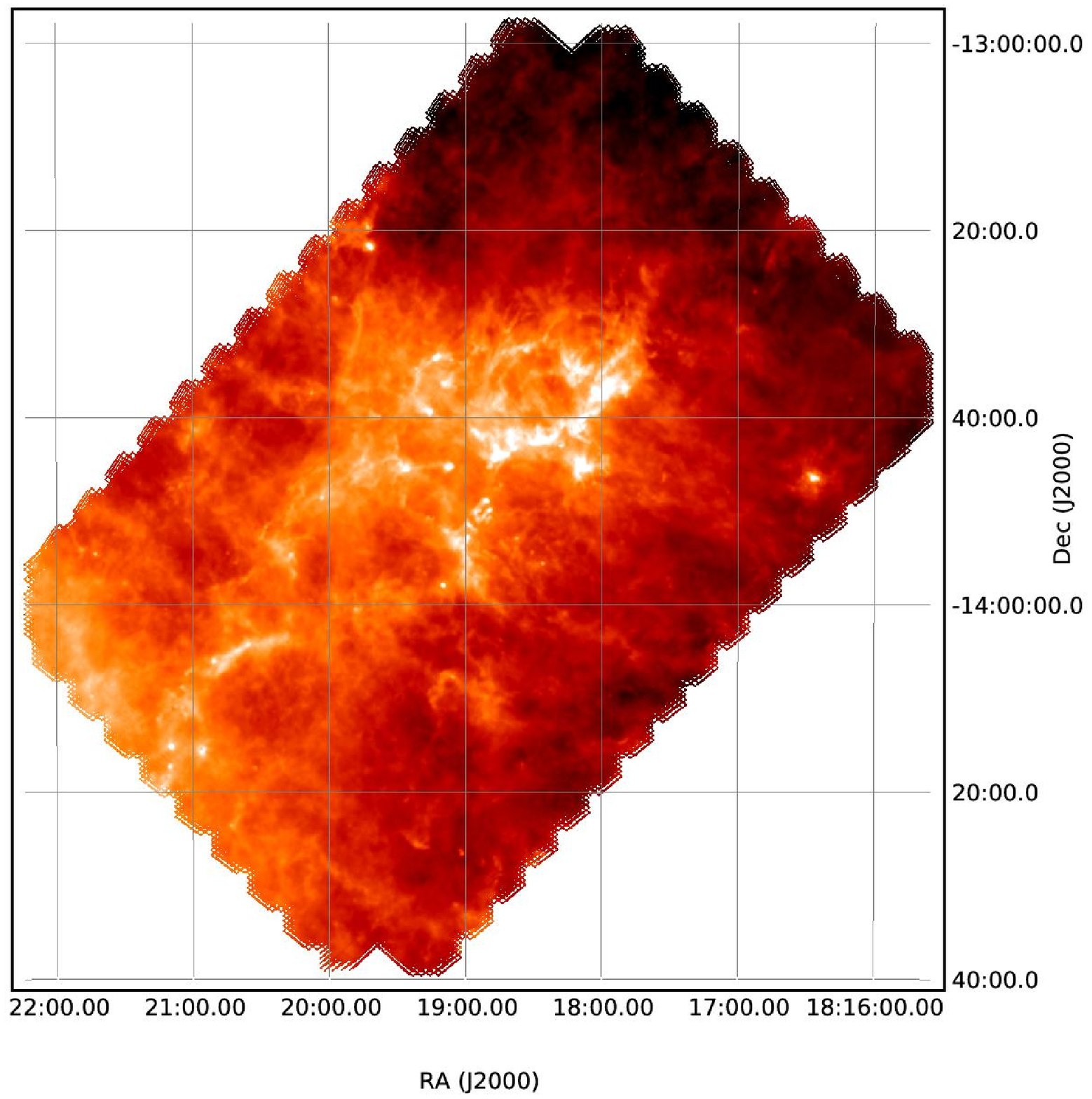}
\includegraphics[width=0.5\textwidth]{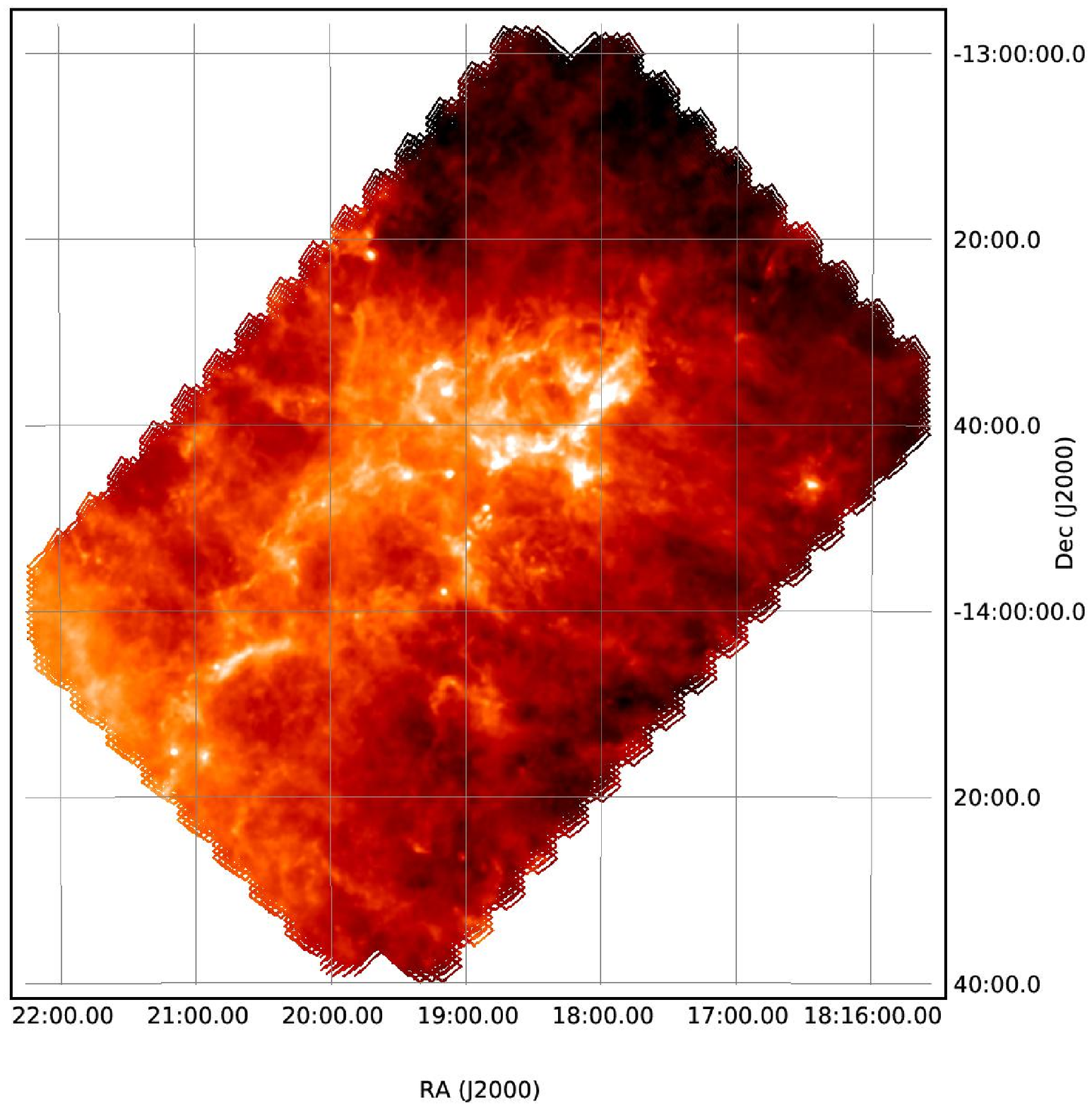}
\begin{minipage}[]{0.48\linewidth}
\includegraphics[width=\textwidth]{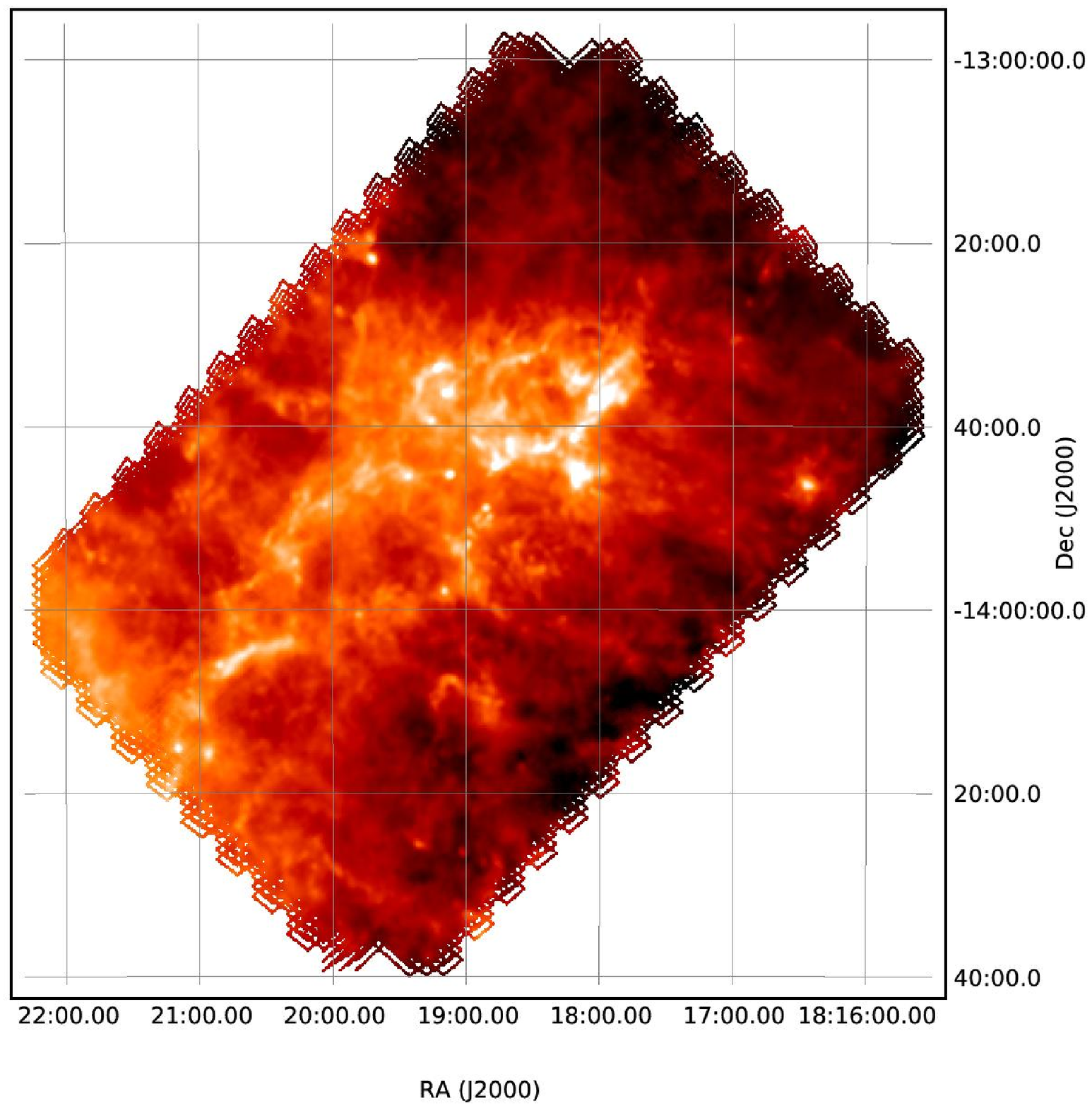} 
\end{minipage}
\hspace{4mm}
\begin{minipage}[]{0.42\linewidth}
\caption{The 5 \herschel\ wavebands of M16, from top to bottom, 70, 160, 250, 350 and 500\,\um. Note also the extended PACS coverage compared with SPIRE (see Section \ref{sec:obs}). The observed level of cirrus noise in M16 is $\sim$ 20\,mJy and 200\,mJy at $\lambda$\,= 70 and 160\,\um, respectively and 2, 1 and 1\,Jy at the SPIRE 250, 350 and 500\,\um\ bands. At shorter wavelengths
only those warmer objects, such as protostars and \hii\ regions are seen. At longer wavelengths \herschel\ detects cold, deeply embedded filaments
and the progenitors of high-mass stars.
\label{fig:hipe:all}}
\end{minipage}
\end{figure*}

\section{Probability distribution functions}\label{sec:pdf}

\begin{figure}
\includegraphics[width=8cm]{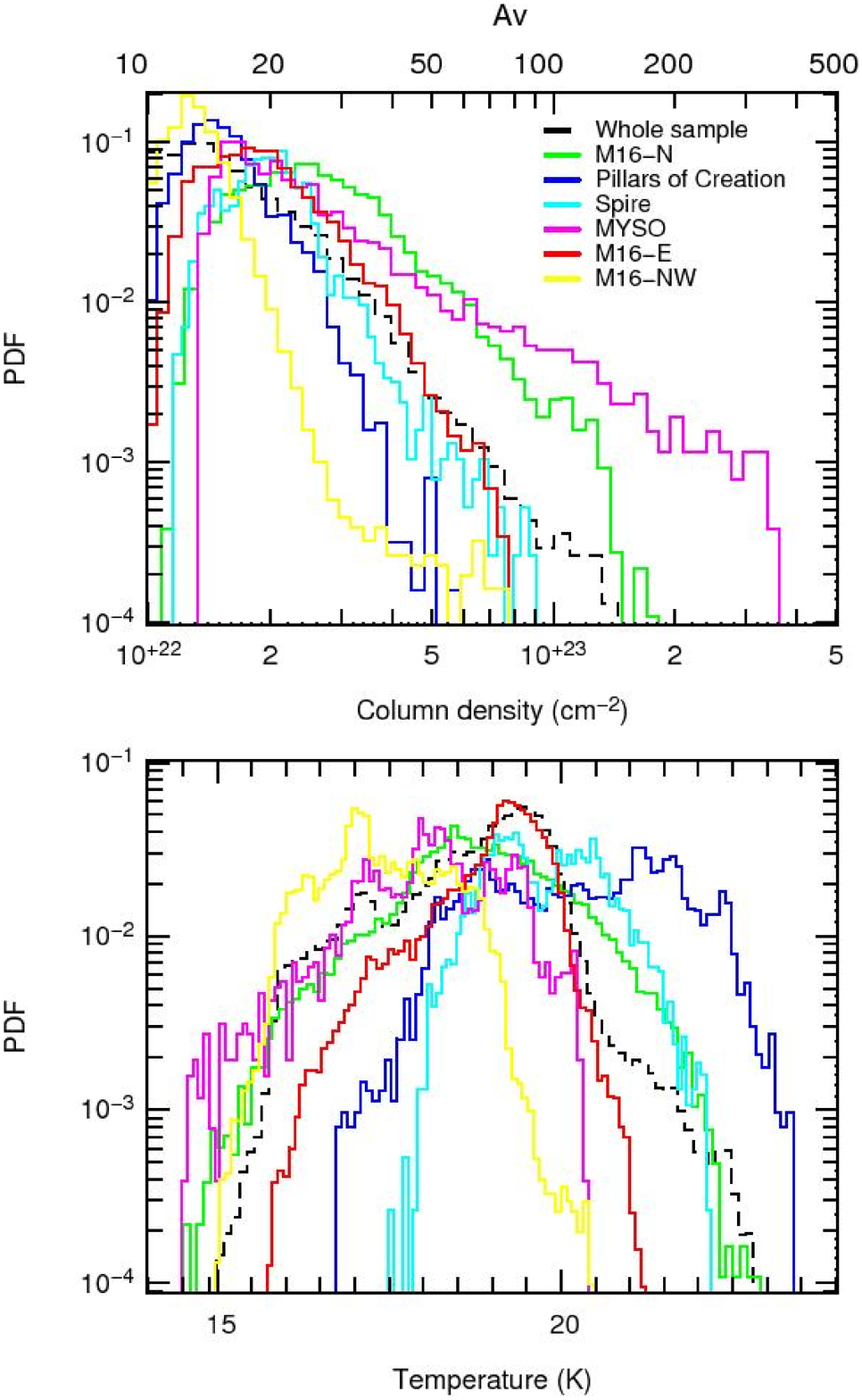}
\caption{PDFs of column density (top) and temperature (bottom) with the sub-regions as defined on Fig. \ref{fig:colden}. Only the high column density material that we are interested in $>$\,10\,mag is shown.  \label{fig:pdf}}
\end{figure}

To examine the dust temperature and column density structure in the entire M16 region probability distribution functions (PDFs) were derived (Fig. \ref{fig:pdf}). 
Column density PDFs have been used by many authors to characterise cloud structure based on star formation activity \citep{kain09, federrath10b}.
The column density in M16 spans almost two orders of magnitude  (5.0\,$\times$\,10$^{21}$ -- 3.7\,$\times$\,10$^{23}$\cmsq), whilst the dust temperature ranges from 14.5\,K to 23.5\,K. The highest temperature in the maps is recorded directly west of the base of the Pillars of Creation and thus south of the NGC\,6611 cluster. 
In order to investigate the variation of PDF profiles within M16, we segregated the complex into six sub-regions (see Fig. \ref{fig:colden}, left), according to the prominent structures or cloud features impacted by the NGC\,6611 cluster. These sub-regions include: the main component of material, corresponding to the Eagle's wings in optical images and the  arch structure (hereafter referred to as the `M16-N' sub-region; green in Fig.~\ref{fig:pdf}), the Pillars of Creation (blue), Spire (cyan), the MYSO detected by \citet[magenta][]{indebetouw07}, the eastern portion of the map (hereafter `M16-E'; red), and the northern periphery (hereafter `M16-NW'; yellow). The MYSO is treated as a separate sub-region due to its high column density tail, to prevent this MYSO from skewing the PDFs of other sub-regions. While the column density PDF of this MYSO should be treated with caution, its inclusion here allows comparison with the other sub-regions with respect to dust temperature.

The column density PDFs  are reliable for the larger regions (e.g., M16-N, M16-E, M16-NW) above \Av\,$>$\,10\,mag\footnote{Where N$_{H_2}$ =  1\,\Av\ $\times$ 10$^{21}$ cm$^{-2}$ mag$^{-1}$ \citep{bohlin78}.}, whilst the smaller regions (e.g., Pillars of Creation, Spire and the MYSO) are reliable only above \Av\,$>$\,20\,mag. We are only concerned with higher column density ($>$\,20\,mag) here.
The M16-N, Spire and M16-E sub-regions have similar column density slopes (around \Av\ $>$\,30\,mag). The M16-N and MYSO sub-region have high column density tails, similar to that seen in the Vela~C Centre-Ridge sub-region \citep{hill11}, and consistent with regulation by self-gravity \citep{bp11}.

The Pillars of Creation sub-region has a column density PDF that resembles a log-normal shape which has been seen in clouds which are initially quiescent and where turbulence is likely the main shaping mechanism \citep{kain09}.
Note, however that feedback effects, such as those from a nearby OB cluster, have yet to be factored into turbulence simulations from which model PDFs are drawn. The suggestion of turbulence here is consistent with the observations of \citet[][$^{12}$CO, $^{13}$CO and C$^{18}$O]{pound98} who found that the Pillars of Creation are not gravitationally bound and are being ripped apart. 
The sub-region hosting the Pillars of Creation (Fig., \ref{fig:pdf}, blue) attains higher dust temperatures than the other sub-regions, including that of the \spitzer\ identified MYSO. This difference shows that the NGC\,6611 cluster has had more of an impact on the cloud temperature than the current star formation process.
The M16-N and Spire sub-regions have similar dust temperature PDF slopes ($>$ 21\,K), which likely reflects their similar distance from the NGC\,6611 cluster, though the M16-N sub-region contains colder gas in the northern part of the region, including the ridge itself.
The M16-NW and M16-E sub-regions span only small ranges in temperature.  The median column density and dust temperature for each sub-region, determined from the PDFs, are given in Table \ref{tab:4band:pdf}. 

It is interesting to note that while the M16-E and Spire sub-regions have similar column density PDF profiles, they display quite different dust temperature PDF profiles. This suggests that they are both a part of the same cloud structure, but that the Spire is closer to and thus more readily externally heated by the NGC\,6611 cluster than M16-E.
The three sub-regions closest to the OB cluster (Pillars of Creation, Spire, MYSO) have similar dust temperature profiles, but quite different column densities.

There is no clear trend in the column density
 PDFs to suggest or represent an effect of the winds or radiation from the NGC\,6611 cluster on the sub-regions.  The median temperature of each region correlates closely with the projected distance from the cluster, suggesting no strongly unbalanced projection effects for the cloud surrounding the NGC\,6611 cluster.

\renewcommand{\thefigure}{C\arabic{figure}}
\setcounter{figure}{0}

\begin{figure*}[!h]
\includegraphics[height=0.28\textwidth]{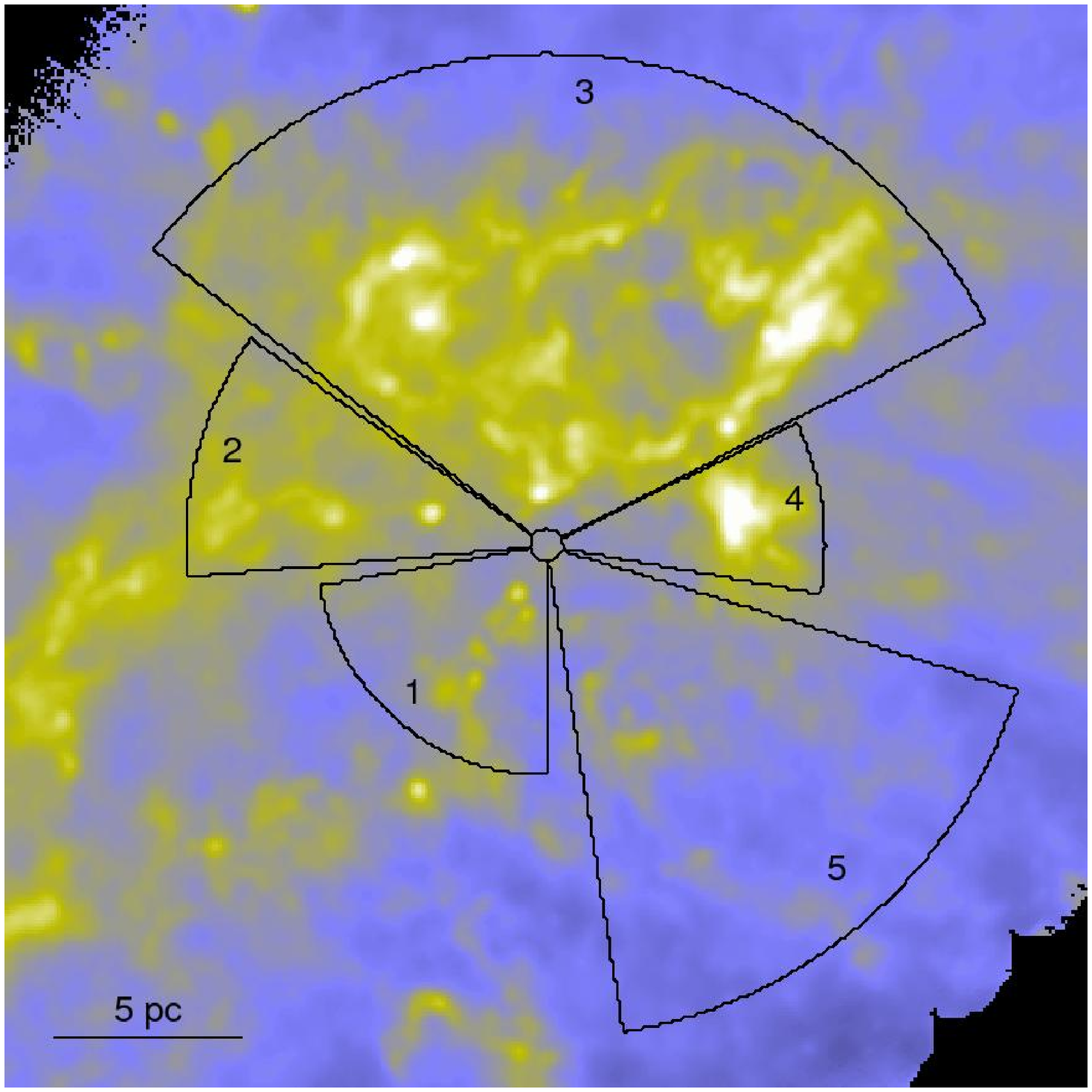}
\hfill
\includegraphics[height=0.28\textwidth]{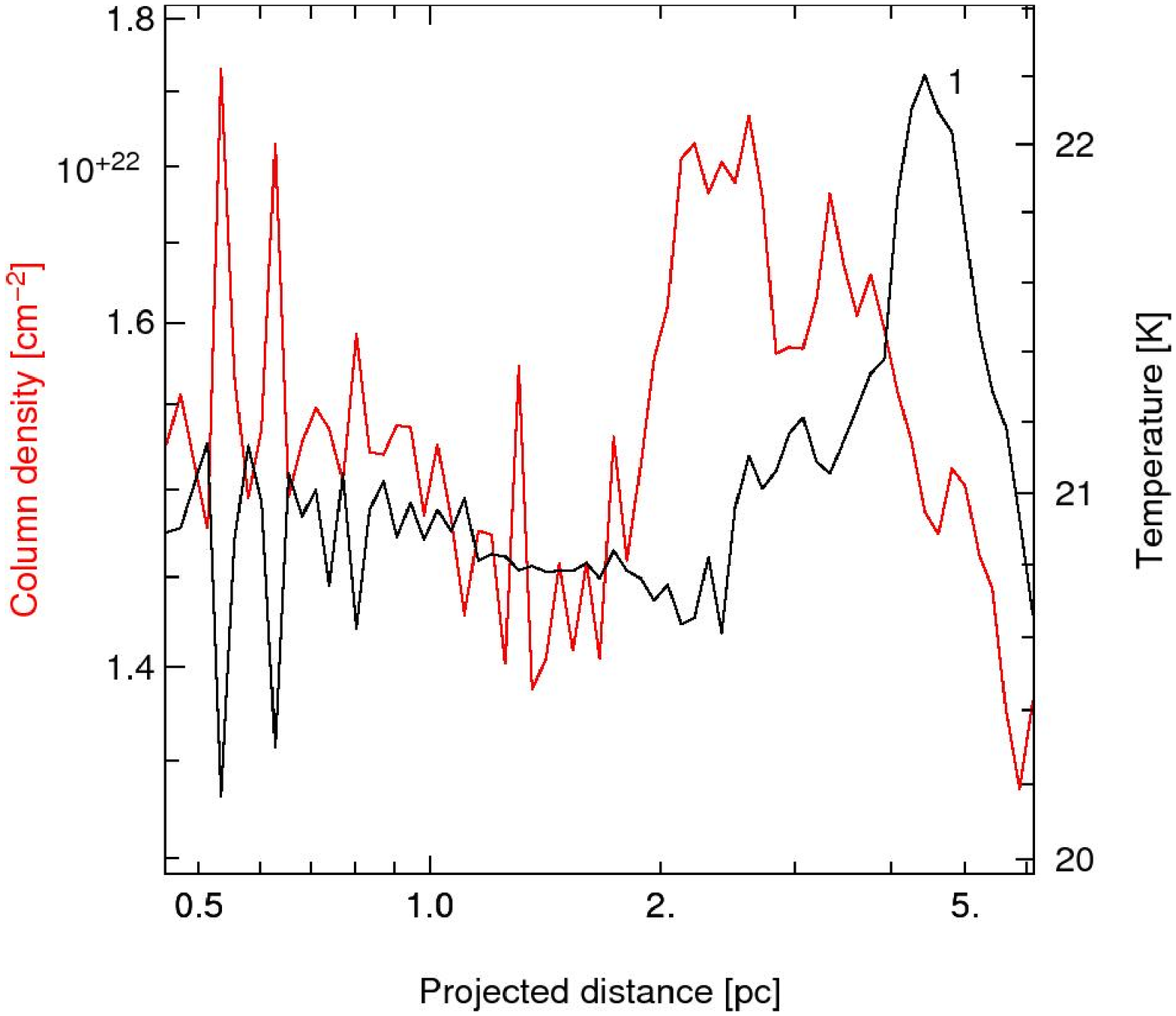}
\hfill
\includegraphics[height=0.28\textwidth]{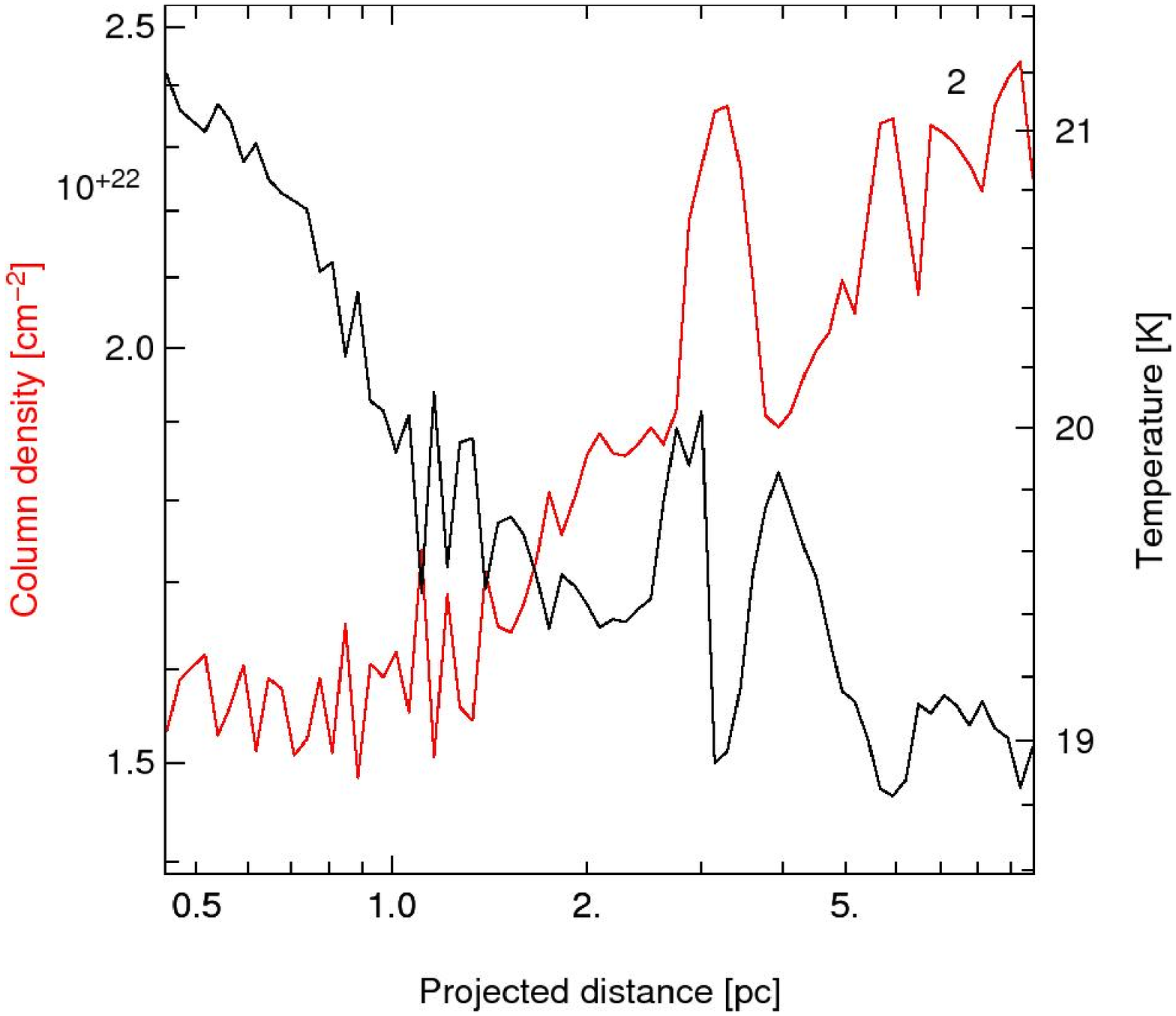}\\
\vfill
\includegraphics[height=0.28\textwidth]{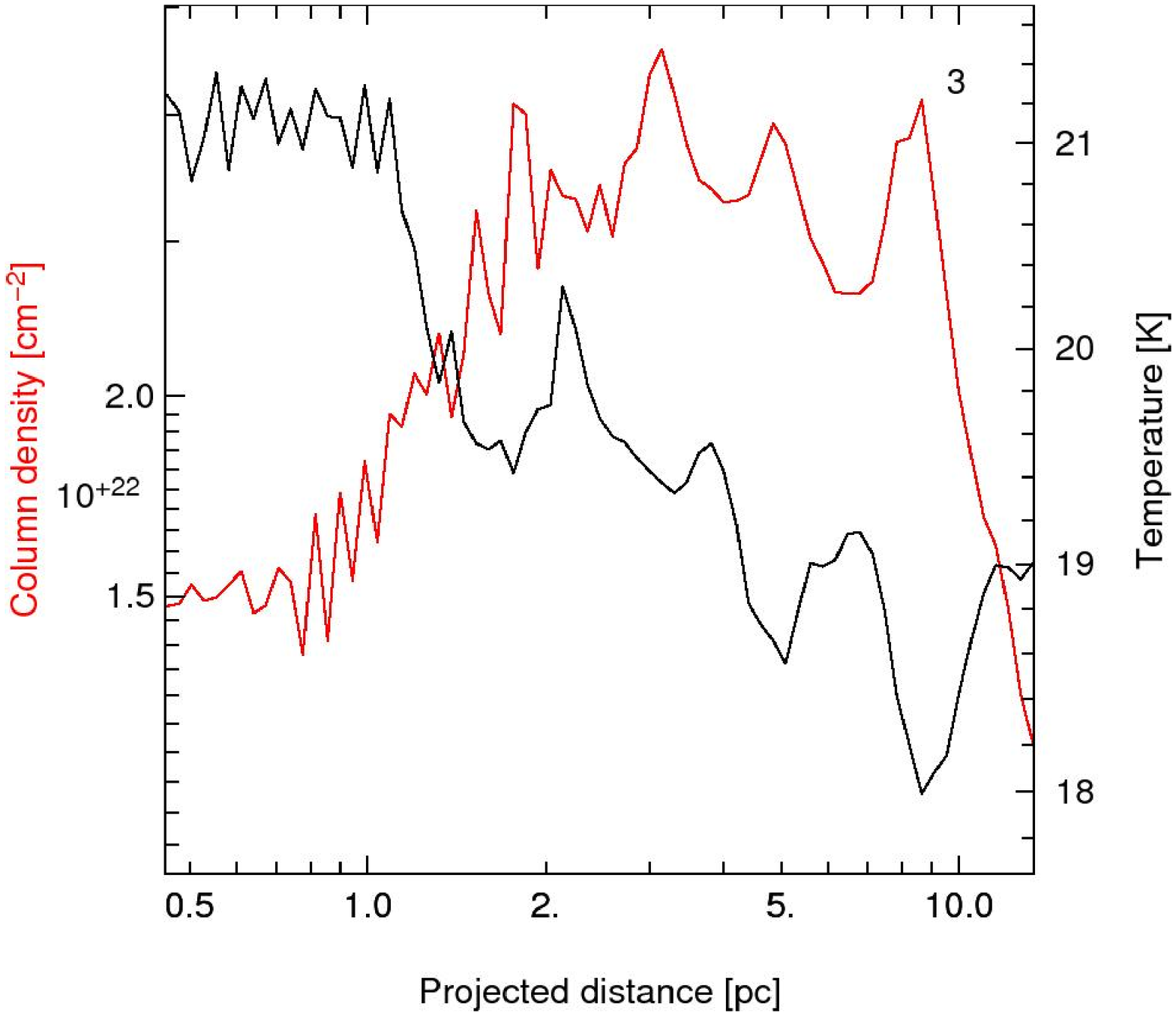}
\hfill
\includegraphics[height=0.28\textwidth]{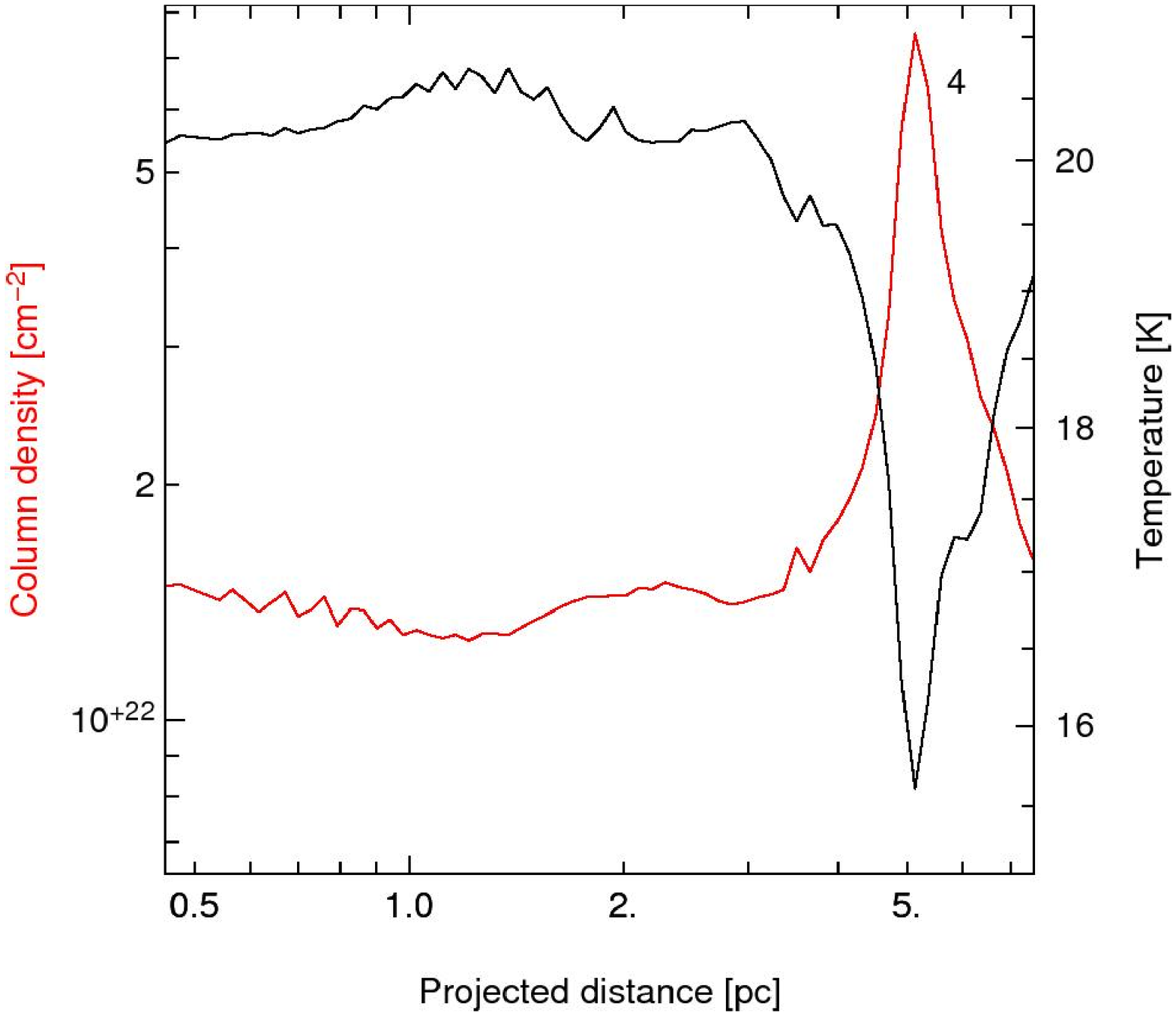}
\hfill
\includegraphics[height=0.28\textwidth]{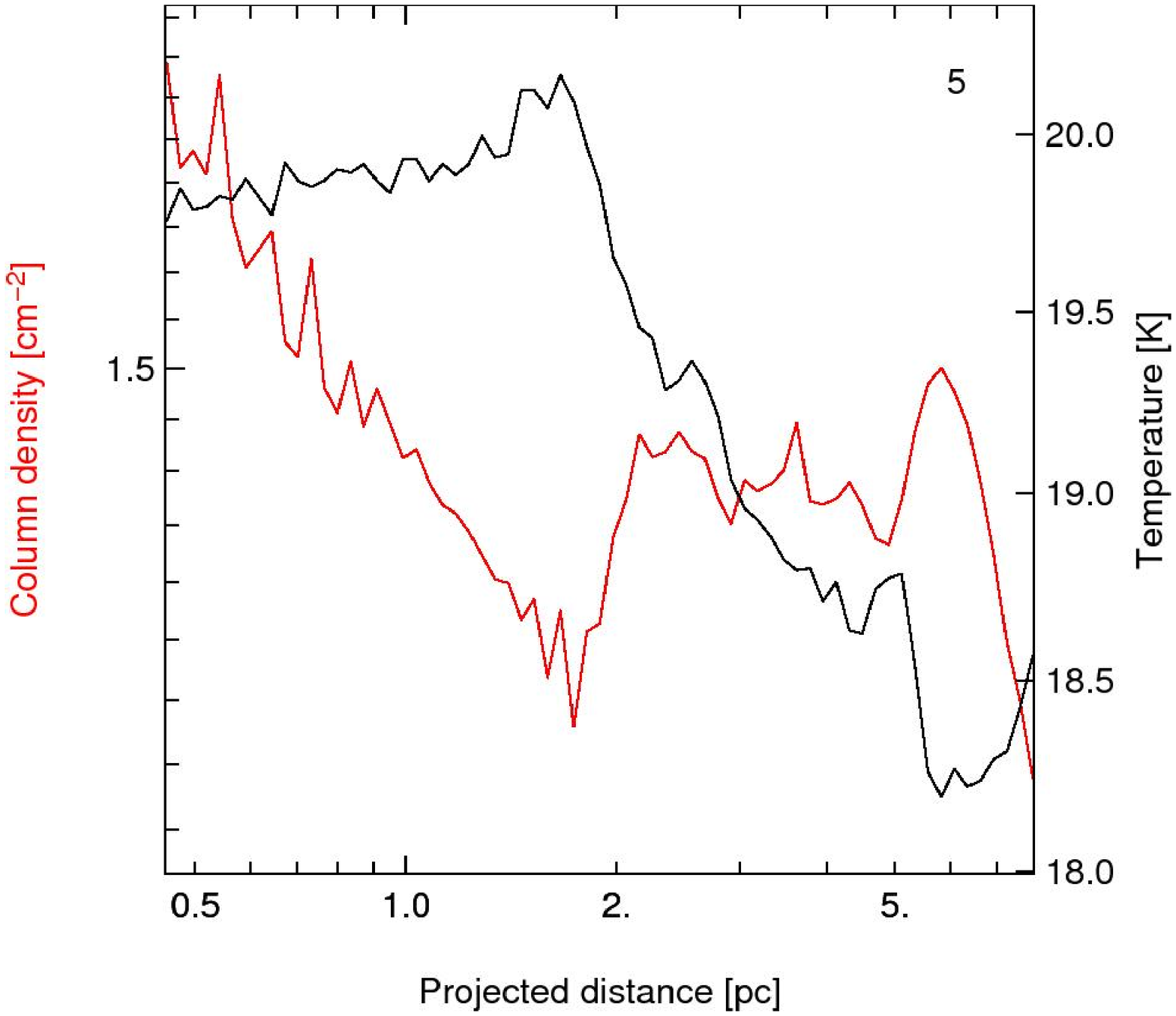}
\caption{Upper left: Column density map of NGC\,6611 with radial segments as indicated. Upper-middle to Lower-right: Dust temperature (red) and column density (green) PA-averaged radial  profiles  in each segment. Segments are arranged clockwise starting with the Pillars of Creation (upper-middle) to the cavity region (lower-right) as identified by numbers. The profiles are the median values with projected distance in each segment to avoid biases by the presence of protostars. The radial lengths of each segment were defined by a drop in column density (as seen in the upper-left panel).  \label{fig:rp:all}}
\end{figure*}

The probability distribution functions (Fig. \ref{fig:pdf}) confirm that the column density in each sub-region in M16 is indeed different, which influences the extent of the heating effect of the NGC\,6611 cluster on each sub-region, as seen in the dust temperature PDFs.
At zero-order and above \Av\,$>$\,15\,mag there is no difference seen in the structure of the cloud with distance from the NGC\,6611 cluster. That is, the column density PDFs of sub-regions near to the cluster, and thus heavily impacted, and those in more remote portions of the map, display similar column density PDF slopes.

It is also worth noting, that though normalised, the Pillars of Creation, Spire and MYSO sub-regions are comprised of a small number of pixels compared with the other sub-regions in the complex, which could thus affect their interpretation.

\section{Detailed radial segments around NGC\,6611}

The radial profiles of more detailed segments (than Fig. \ref{fig:rp}) around the NGC\,6611 cluster, including the Pillars of Creation, can be found in Fig. \ref{fig:rp:all}.

\end{appendix}
\end{document}